\documentclass[fleqn,usenatbib]{mnras}

\usepackage{color}
\usepackage{graphicx}
\usepackage{footnote}
\usepackage{tablefootnote}
\usepackage[caption=false]{subfig}
\usepackage{amsmath}
\usepackage[authoryear]{natbib}
\bibpunct{(}{)}{;}{a}{}{,}
\def\mydoubleq#1{``#1''}

\title[Molecules at Low Frequencies ]{A First Look for Molecules between 103 and 133\,MHz using the Murchison Widefield Array}

\author[Tremblay et al.]{Chenoa D. Tremblay$^{1}$\thanks{E-mail:
chenoa.tremblay@postgrad.curtin.edu.au}, Natasha Hurley-Walker$^{1}$, Maria Cunningham$^{3}$, \and Paul A. Jones$^{3}$, Paul J. Hancock$^{1, 2}$, Randall Wayth$^{1}$ and Christopher H. Jordan$^{1,2}$
\\
$^{1}$International Centre for Radio Astronomy Research, Curtin University, GPO Box U1987, Perth WA 6845, Australia\\
$^{2}$ARC Centre of Excellence for All-sky Astrophysics (CAASTRO)\\
$^{3}$School of Physics, University of New South Wales, Sydney, NSW 2052, Australia\\
}

\date{Accepted XXX. Received YYY; in original form ZZZ}

\pubyear{2015}

\begin{document}
\label{firstpage}
\pagerange{\pageref{firstpage}--\pageref{lastpage}}
\maketitle

\begin{abstract}
%Historically, astrochemistry has been studied at frequencies greater than 80\,GHz, but line confusion from multiple transitions can make identification of new molecules difficult.  
We detail and present results from a pilot study to assess the feasibility of detecting molecular lines at low radio frequencies. We observed a 400 square degree region centred on the Galactic Centre with the Murchison Widefield Array (MWA) between 103 and 133\,MHz targeting 28 known molecular species that have significant transitions. The results of this survey yield tentative detections of nitric oxide (NO) and the mercapto radical (SH). Both of these molecules appear to be associated with evolved stars.  %In this paper, we discuss the method developed for this work and the importance of this work. 

\end{abstract}


%%%%%%%%%%%%%%%%%%%%%%%%%%%%%%%%%%%%%%%%%%%%%%%%%%

%%%%%%%%%%%%%%%%% BODY OF PAPER %%%%%%%%%%%%%%%%%%

\section{Introduction}
In the last five years, new radio telescopes have been commissioned for use in the frequency range of 30 to 300\,MHz, including the Long Wavelength Array (LWA; \citealt{Taylor}), the Low-Frequency Array (LOFAR; \citealt{vanHaarlem}) and the Murchison Widefield Array (MWA; \citealt{Tingay13, Lonsdale}). The capabilities of this new generation of radio telescopes, as well as upgraded facilities such as the Karl G. Jansky Very Large Array (JVLA; \citealt{JVLA}) and Giant Metrewave Radio Telescope (uGMRT; \citealt{uGMRT}), allow us to revisit low radio frequencies to study spectral lines.  

Radio spectral line emission from within our Galaxy is a widely-used probe of astrophysics \citep{Herbst09}.  Comparison of different spectral lines, and therefore different molecules, can be used as a probe of many interstellar processes including the chemical and physical evolution of stars \citep{2014FaDi..168....9V} and planets \citep{Cosmovici79}. Many of the recent large molecular line surveys in the radio, infrared, and submillimetre, such as MALT 45 \citep{Jordan15}, MALT 90 \citep{Jackson}, Herschel$/$Hi-GAL follow-up \nocite{Olmi2, Olmi} (e.g. Olmi et al. 2014, 2015) and HOPS \citep{Walsh11} were motivated by studying and characterising molecular clouds, high mass stars, and their environments.  

One of the challenges faced by sensitive spectral line surveys at high frequencies is line confusion of multiple transitions, making identification of new molecules difficult. For example, the IRAM 30\,m survey of the Orion region between 80 and 281\,GHz, is significantly hindered by known prominent spectral lines when searching for rarer molecules such as methyl formate (CH$_{3}$OCHO; \citealt{Tercero15}).
As we observe in frequencies down towards 100\,MHz, these transition spread out, making them less confused.

The formation pathways for many molecules, especially for complex organics, are still unknown. Recent articles suggest that observations at the lower frequencies may provide key information by observing long chain molecules and low energy transitions of simple diatomics that are not easily identified with observations at higher frequency (e.g. \citealt{Codella14, Danilovich}). This was demonstrated recently in the search for propylene oxide (c-CH$_{3}$C$_{2}$H$_{3}$O) where previous attempts to find the molecule at 99\,GHz \citep{Cunningham07} failed but more recent results at 12.1\,GHz yielded detections of three transitions \citep{2016Sci...352.1449M}.

Low-frequency observations may also be beneficial to study the formation mechanisms of high-mass stars through the detections of molecules typically hindered by line confusion in higher frequencies \citep{Codella14}. \cite{Isella15} suggest that the ionised hydrogen ({\sc Hii}) regions around high mass stars may be optically thick, hiding the central
molecular emission of high mass protostars, at frequencies less than 10\,GHz. \cite{HindsonL} found that the {\sc Hii} regions are optically thin until around 200\,MHz, suggesting that observations between 1\,GHz and 200\,MHz could provide contributions to understanding the chemical evolution of high-mass stars and thus their formation in regions where the ionised hydrogen is uniform. Lower frequencies could still contribute in early stages of formation before the star ignites and when the gas is clumpy.

Sulfur is one of the most abundant elements in the universe.  However, there is an apparent lack in molecular form in the interstellar medium. 
\cite{Gorai} review the current issues regarding the missing interstellar sulfur-containing molecules during their research of the formation mechanisms of thiols\footnote{The sulfur analogue of alcohols}. Currently, observations have only accounted for one-quarter of the total interstellar sulfur thought to be associated with molecules, which may be due to molecular instability at higher temperatures \citep{Millar}.

Motivated by the possibility of detecting molecular spectral lines at radio frequencies that are not normally accessible (due to line confusion, instrumentation limitations or radio-frequency interference), we have used the MWA to perform a pilot survey around the Galactic centre region.  Within this region we would expect to detect molecular lines within evolved stars and molecular clouds. However, with the dominance of synchrotron radiation at these frequencies and the lack of good prediction models, it  is difficult to know what to expect. A primary science goal of the MWA is the search for redshifted \textsc{Hi} 21\,cm emission from the Epoch of Reionisation (EoR) \citep{Bowman}. The design requirements for the EoR experiment make the MWA well suited for large-area blind spectral line surveys.

\begin{figure*}
	 \includegraphics[width=1\textwidth]{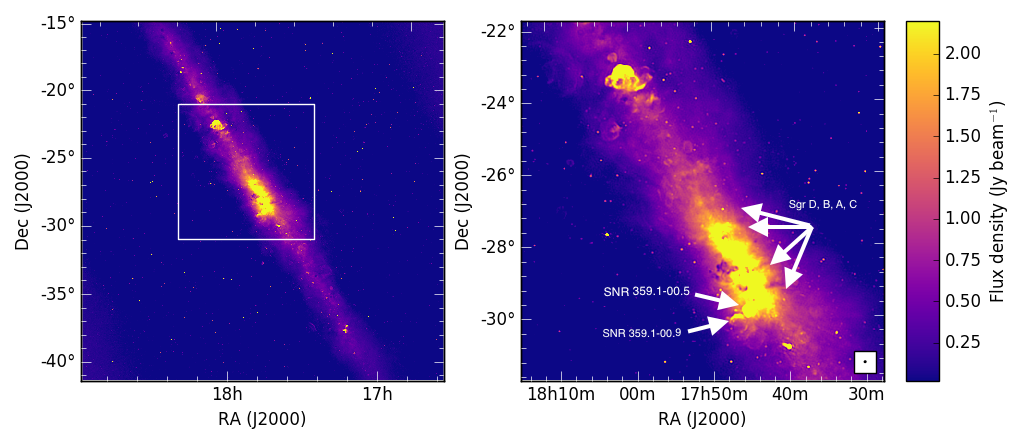}
	\caption{Continuum image of the Galactic Centre observed by the MWA across the 30\,MHz of bandwidth with a central frequency of 119.7 MHz. The left hand image represents the full region blindly searched for molecular signatures (see section \S4.2).  The right-hand image is a zoomed in region around the Galactic Centre showing that the supernova remnants SNR~$359.01-00.5$ and SNR~$359.0-00.9$ are clearly resolved.  The star-forming regions of Sagittarius~B, Sagittarius~C, Sagittarius~D and Sagittarius~A are unresolved in our observations. }
	\label{GC}
\end{figure*}

The primary goal of this work was to assess the feasibility of using a new, multi-purpose, wide-field, low-frequency radio telescope, such as the MWA, to study molecular spectral lines and to demonstrate a pipeline for processing MWA data for such an analysis.

\begin{figure}
	 \includegraphics[width=0.49\textwidth]{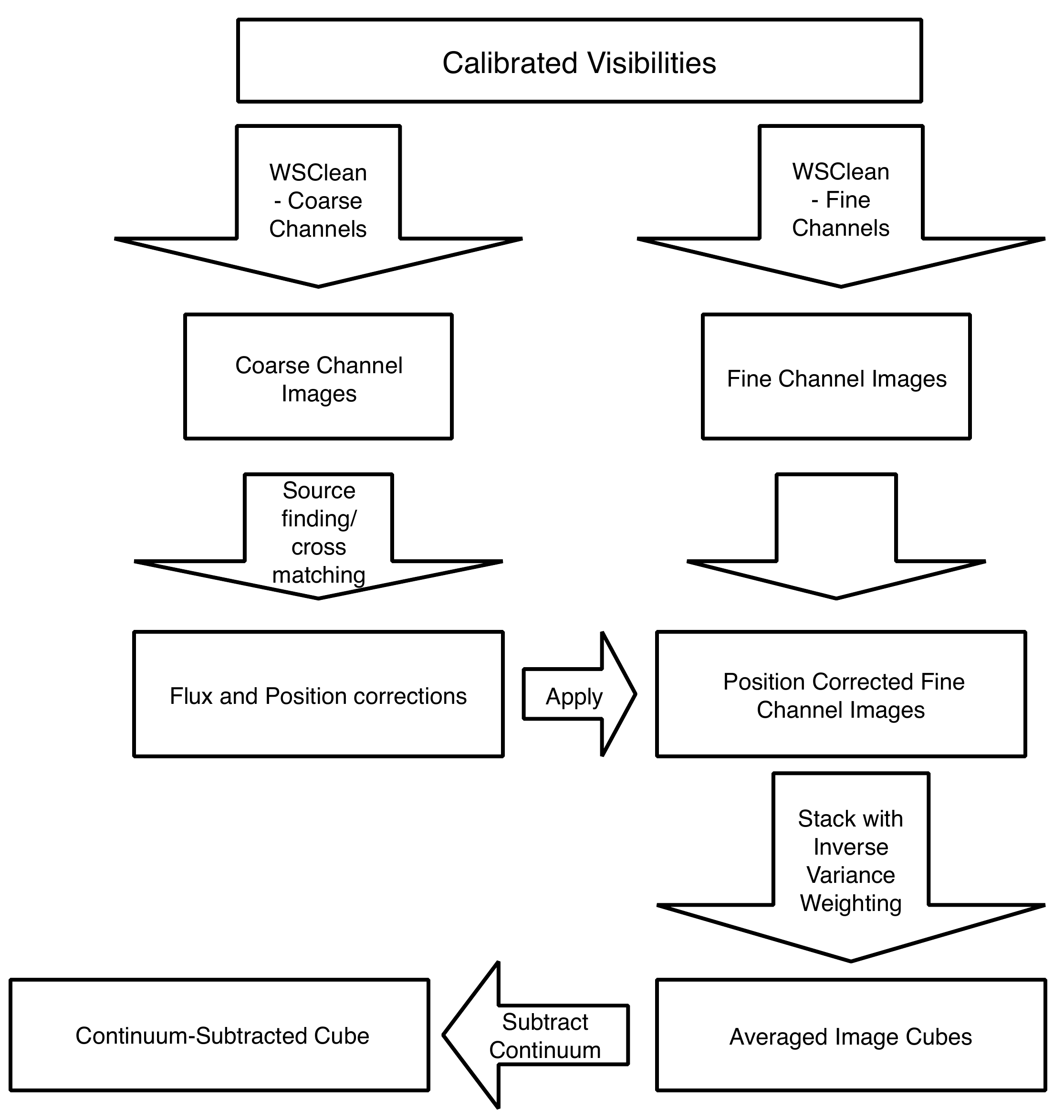}
	\caption{Summary of the pipeline designed to create calibrated time-averaged spectral image cubes with the MWA.}
	\label{pip}
\end{figure}

%%%%%%%%%%%%%%%%%%%%%%%%%%%%%
%%%%%%%%%%%%%%%%%%%%%%%%%%%%%
\section{Observations \& Data Reduction} \label{Obs}

The region under study, shown in Figure~\ref{GC}, is 400 square degrees centred on 17$^{h}$45$^{m}$40$^{s}$ $-$29$^{\circ}$00$^{\prime}$ 28$^{\prime\prime}$ (J2000).  

The observations were performed using the MWA, which is located at the Murchison Radio-astronomy Observatory in Western Australia, on 25 July 2014 and 27 July 2016 for 120 minutes each night.
A detailed description of the MWA, including its capabilities and specifications, are found in \cite{Tingay13}.  However, we provide a short outline and details pertinent to this work.
The MWA consists of 128 antenna \mydoubleq{tiles}, each containing 16 dual-polarisation dipole antennas, with a maximum baseline length of 3\,km. At 120\,MHz, the FWHM of the primary beam (field-of-view) and synthesised beam (resolution) are 30\,degrees and 3.2\,arc minutes, respectively. 

For each of the two nights, the observations were completed in 24 five-minute snapshots using a standard MWA observing procedure. At the beginning of each five-minute snapshot, the antenna tiles are re-pointed to track the field over the night.

Data were collected in a 30.72\,MHz contiguous band, centred at 119.7\,MHz.
The MWA uses a two-stage filterbank to channelise the signal.
The output of the first filterbank consists of $24\times1.28$\,MHz ``coarse'' channels; each coarse channel is further subdivided into $128\times10$\,kHz ``fine'' channels, resulting in $3072\,(10$\,kHz-wide) spectral channels.  The 10\,kHz channel width corresponds to a velocity resolution of 26\,km\,s$^{-1}$.   

The edges of each coarse channel suffer from aliasing in the first stage filterbank, hence the 14 fine channels on both edges of each coarse channel were flagged. The central fine channel of each coarse channel contains the (non-zero) DC component of the polyphase filterbank \citep{Thiagaraj} and was also flagged.  Additional flagging for radio-frequency interference was performed in each observation using standard MWA tools that are based on \textsc{AOFlagger} \citep{Offringa10,Offringa12}.  This program is designed to find peaks in time and frequency that are likely to be the result of radio frequency interference (RFI).  

In summary, the central 100 fine channels of each coarse channel were imaged. This resulted in 78{\%} of the bandpass being imaged. 

\subsection{Calibration \& Imaging} \label{Process}

The MWA tool {\sc Cotter} \citep{Offringa15} was used to apply the flags and set the phase centre for every observation to 17$^{h}$45$^{m}$40$^{s}$ $-$29$^{\circ}$00$^{\prime}$ 28$^{\prime\prime}$ (J2000).  

The bright radio source Hercules~A was observed for two minutes each night and was used for initial bandpass calibration, using the {\sc Mithcal} calibration tool described in \cite{Offringa16}.  Radio source 3C353 also lies within the field-of-view of the observation centred on Hercules~A, hence we followed the approach adopted by \cite{HW14} to incorporate 3C353 using self calibration.  The revised bandpass solution was then applied to each of the five-minute observations of the Galactic Centre.

As this was the first attempt to analyse MWA observations for spectral lines, we designed a novel method to create searchable image cubes; the pipeline is summarised in Figure \ref{pip}.  

In widefield interferometric imaging with a non-coplanar array, the visibility ($u,v,w$) data can no longer be related to the sky ($l,m$) by a simple 2-D Fourier Transform, because the values of $w$ are no longer negligible \citep{Cornwell05}. WSClean \citep{Offringa14},  uses $w$-stacking to perform a computationally-efficient transformation of visibility data into sky images. This software was used to simultaneously clean and image the fine channels associated with each coarse channel by searching for the clean components in a multi-frequency synthesis image and then subtracting the components from each individual fine channel of the visibilities. We used a Briggs weighting \citep{Briggs} of \mydoubleq{$-1$} for the gridded visibility data to compromise between image resolution and sensitivity.

For each of the 24 coarse channels, 100 of the 10kHz fine channels were imaged  to ensure that channels that may be affected by aliasing have been removed.  The 10\,kHz channel images, made using 4.2 pixels per synthesised beam FWHM, were combined to produce a single image cube which was then converted into a {\sc miriad} \citep{Sault} file format. In {\sc miriad}, all of the five-minute snapshot images were time-averaged together using inverse variance weighting,  after flux density calibration (\S 2.1.1) and ionospheric correction (\S 2.1.2), to form a single image cube for each coarse channel, representing 235\,minutes of integration time.  

As shown in Figure \ref{pip}, a continuum image of each coarse channel is used to derive the primary beam model, check the flux density calibration and determine the ionospheric correction values that are applied to each of the fine channel continuum images.

\begin{figure}
	 \includegraphics[width=0.51\textwidth]{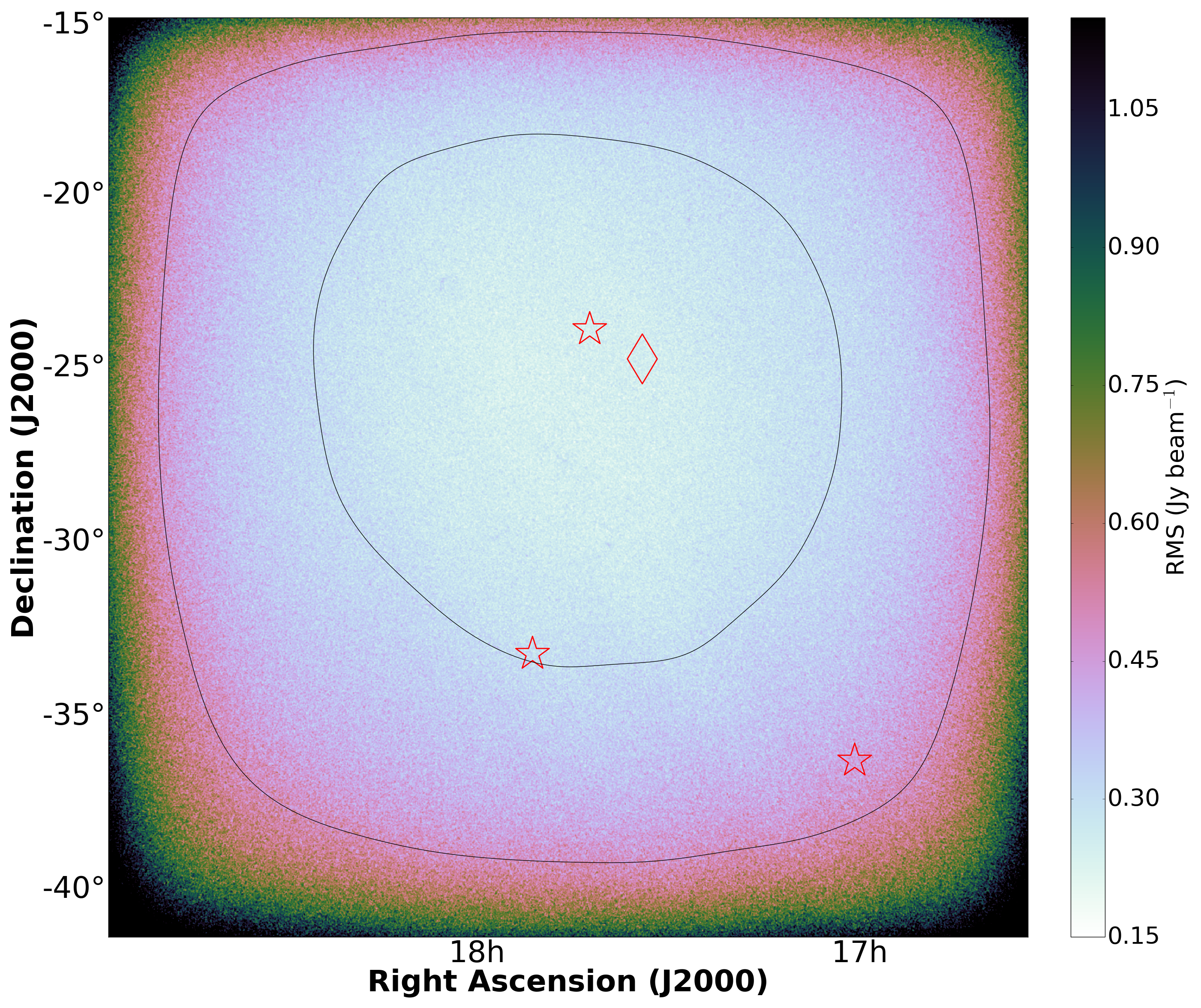}
	\caption{Map of the RMS following the shape of the primary beam across the searched field-of-view.  The cube of the fourth coarse channel, at 107\,MHz, is shown in this figure, as a representation of the behaviour of the entire dataset. The stars on the image represent the positions of the nitric oxide tentative detections and the diamond represents the position of the mercapto radical tentative detection.  See \S 4 for additional details.  The contours are at the levels of 0.3 and 0.5\,~Jy\,~beam$^{-1}$}
	\label{rms}
\end{figure}

Observations at these low frequencies are sky-noise-dominated. The measured noise within the continuum subtracted image, after primary beam correction, is approximately 150\,mJy\,beam$^{-1}$ for most image cubes at the pointing centre and lower than the 300\,mJy\,beam$^{-1}$ theoretical RMS derived from the background sky temperature and the system temperature reported in \cite{Tingay13}. However, the RMS noise increases to approximately 300\,mJy\,beam$^{-1}$ six degrees from the Galactic Centre, following the shape of the primary beam as shown in Figure \ref{rms}.  The estimated sky temperature, through a scaling of the Haslam model \citep{Remazeilles}, for the Galactic Centre was T$_{\rm sky} =3582$\,K, which was a factor of ten greater than the nominal T$_{\rm sky}$, due to the bright diffuse synchrotron emission from this part of the Galaxy.  

\subsubsection{Flux Density Calibration}
The primary beam is the sensitivity pattern of the individual receiving elements (tiles for the MWA) and is direction and frequency dependent. The dipoles are not uniformly sensitive to the incoming radiation from all directions, reducing the apparent flux density of sources located further away from the pointing centre. Therefore, we must correct for the changing brightness distribution on the sky by dividing the fine channel continuum image by the beam model derived from each coarse channel for each observation.  The flux densities of observed sources are then compared to those published in the Molonglo Reference Catalogue (MRC; \citealt{Large}), as the 408\,MHz catalogue is close to our observing frequencies and it has sources within $\mid$b$\mid$ $>$ 3.  The values for each cross-matched point source in the MRC were scaled down to our frequency using a spectral index of $-0.83$ \citep{GLEAM}.  The flux densities were found to have a flux density scale difference of 3\% over 25 sources, so no direction-dependent calibration was required.

\subsubsection{Ionospheric Correction}
Each five-minute observation over the two epochs experienced slightly different phase distortions due to changing ionospheric conditions. This manifested as small ($\approx20$--$40$\,arc seconds) direction-dependent shift to the positions of sources in every observation. Using the software  {\sc aegean} \citep{Hancock12}, the source positions within each coarse channel continuum image was determined.  These positions were compared to those reported in the MRC and the mean shift was applied to each image cube, before the observations were averaged together (\S 2.1).    

After corrections were made and the image cubes were averaged together, the residual shift in source position, compared to MRC, is 1\,arc second in right ascension (RA) and $-$2\,arc seconds in declination (Dec).   The ratio of the peak flux density and the integrated flux density was calculated to determine the amount of blurring from residual uncorrected ionospheric shifts and found to be 1.10$\pm$0.11.   Based on the standard deviation of the position offsets of the sources from MRC, the astrometric precision for a typical source position is 18\,arc seconds with the dominant error being uncorrected residual shifts within each observation.  

\subsubsection{Continuum Subtraction}
The MWA is made up of dipole antennas that observe a wide field of view, so traditional methods of continuum subtraction in $(u,v,w)$ space could not be used.  Early attempts to use the {\sc casa} command {\sc uvcontsub} to subtract the continuum using a first order polynomial, resulted in only removing 71 of the continuum and created severe smearing of sources away from the phase centre.  These were likely due to the inability of {\sc uvcontsub} to correct for the $w$-terms. Therefore, the continuum was subtracted in the image domain. 

We followed a method similar to that which was developed for and used in the Southern Parkes Large-Area Survey in
Hydroxyl (SPLASH; \citealt{Dawson}) to subtract the continuum and residual bandpass.  A smoothed image cube was created by binning together the spectral signal in sets of  15 fine channels, for each pixel position, and subtracting the new smoothed image cube from the continuum image cube.

%%%%%%%%%%%%%%%%%%%%%%%%%%%%%%%%%%%%%%%%%%%
%%%%%%%%%%%%%%%%%%%%%%%%%%%%%%%%%%%%%%%%
\section{Line Search \& Identification} \label{Linesearch}

\subsection{Survey Strategy}  
In this survey, we preformed two different search strategies to find spectral lines within these image cubes.  First, we completed a targeted search by looking for lines we thought were most likely to be detected based on the line strengths, quantum numbers, and energies listed within the databases Cologne Database for Molecular Spectroscopy (CDMS) \citep{Muller}, Spectral Line Atlas of Interstellar Molecules (SLAIM) (Splatalogue\footnote{www.splatalogue.net}), Jet Propulsion Laboratory (JPL) \citep{Pickett} and Top Model (\S 4.1).  Second, we completed a blind search over 1920 of the 2400 imaged spectral channels and looked for lines brighter than the 6$\sigma$ limit of the local RMS within the image cube  in both emission and absorption (\S 4.2).  The Galactic Centre is of particular interest in the blind search as the Galactic centre offers a stimulated environment with increased gas densities, temperatures \citep{Menten} and possibly cosmic rays \citep{Chambers}. These additional factors may be responsible for the fact that transitions from the largest molecules have only been detected toward the Galactic centre \citep{Jones12}.  
 
 \subsection{Line Search}
As this is the first survey of molecular lines at 100\,MHz and many of the channels are affected by aliasing (\S2.1), a conservative search approach was used. A limit of 6$\sigma$ of the local RMS in the final image cube for the blind search and 5$\sigma$ for the targeted search, was set for the molecular line search in both emission and absorption. To perform the search, an RMS map of each averaged image cube, each representing one of the 24 coarse channels, was created.  The RMS map was divided by the continuum-subtracted image cube to create a signal-to-noise (SNR) data cube. The SNR cube was converted into a {\sc miriad} \citep{Sault} file format and a map of the peak intensity of each pixel was created by fitting a three-point quadratic to every three adjacent fine channels in each pixel position (see {\sc miriad} documentation on moment map \mydoubleq{-2} for additional details).\footnote{https://www.cfa.harvard.edu/sma/miriad/manuals/ATNFuserguide\_US.pdf}  The software {\sc aegean} \citep{Hancock12} was used to look for peaks over 6 in each of the 24 coarse channel maps  for the blind search and for peaks over 5 in two of the coarse channel maps for the targeted search.  

Each potential spectral line found by {\sc aegean} was further investigated by observing the location in the continuum subtracted image cube. For any peak associated with a known molecular line, the data were then cross-matched with objects in SIMBAD\footnote{http://simbad.u-strasbg.fr/simbad/sim-fcoo} lying within three arc minutes of the centre pixel, the size of the FWHM of the synthesised beam.

%%%%%%%%%%%%%%%%%%%%%%%%%%%%%%%%
%%%%%%%%%%%%%%%%%%%%%%%%%%%%%%%%
\section{Results}
\label{Results}

\subsection{Targeted Search}

Based on line strengths and energies, given in Tables~\ref{SH}, \ref{NO} and \ref{Molecules}, we expect the most likely detections to be the mercapto radical (SH) and nitric oxide (NO).  Nitric oxide has a significant number of known low energy transitions at the frequencies of 103$-$133\,MHz and was determined by \cite{Gerin} to have abundances similar to C$^{18}$O.  The ground-state hyperfine-splitting transitions of SH lie in the range 100-123 MHz. 

These two molecules have been detected in other transitions towards astrophysical objects. However, we expect some of the excitation mechanisms for these low frequency detections to be different.  Many of the molecular transitions detected at higher frequencies are of shocked or hot gas. The expectation is that at low frequencies the transitions would primarily be emitted from cold gas.

Based on a confirmed Gaussian noise distribution, at a 5\,$\sigma$ level, we would expect 1 false detection per image cube.  However, using the source density described in \S4.2, we would expect 0.003 channels to have a signal that correlates with a known molecular line and a known object.  

We have tentatively detected the Mercapto radical (\S 4.1.1) and nitric oxide (\S 4.1.2).  Both of these detections appear to be associated with evolved stars which are plausible chemical environments for these molecules.  The chances of these peaks being the result of noise, instead of molecular detections is 0.03, using similar source densities and statistics as described in \S4.2.  

\subsubsection{Tentative Detection of Mercapto radical (SH)}
 
Diatomic hydrides are the simplest interstellar molecules and may provide key information about the interstellar medium (ISM); the reactions that lead to their production in interstellar environments are characterised by low activation energies and high critical densities \citep{Bruderer, Godard12}. The lowest energy hyperfine transitions of SH emit at 100.30, 111.49, 111.55 and 122.73\,MHz, all within the range of the MWA (Table \ref{SH}).  The transition at 100.30\,MHz  is outside our observing band.  The lines at 111.49 and 111.55\,MHz are the highest intensity transitions representing $^{2}\Pi_{3/2} J=~\frac{3}{2}$, F~=1$^{+}$-~1$^{-}$ and $^{2}\Pi_{3/2}~J=\frac{3}{2}$,~F=~2~$^{+}~$--2$~^{-}$ respectively, with energy at the upper excitation level of 5$\times$10$^{-4}$\,K.  

We have tentatively detected the $^{2}\Pi_{3/2}~J=\frac{3}{2}$,~F=~2~$^{+}$~-~-~2~$^{-}$ SH transition at 111.56\,MHz around the evolved star 2MASS~J$17360840-2533343$ (Figure~\ref{SH}).  This transition represents the most intense line and so the most likely to be detected. 

The SH hyperfine transitions are analogues to those of the hydroxyl radical (OH) at 1.6\,GHz \citep{Dawson}.  The OH transitions are well-known strong masers that originate from star-forming regions and evolved stars. Therefore, it might be expected that the SH analogue transitions might also exhibit maser activity. This would significantly increase their detectability, although predicting the maser gain for such transitions is difficult.  

If we assume similar thermal emission properties, we can use the known abundance information about OH to predict the flux of SH in these observations. The sensitivity calculation uses a standard ratio of the relative abundance of sulfur compared to oxygen, which is 1/21 \citep{Wilson} and the detected concentration of OH in the Galactic Centre with SPLASH \citep{Dawson}. With this, we estimate SH should show absorption of about 0.3 and 0.5\,Jy for the 111.49 and 111.55\,MHz transitions respectively.  This  intensity is around the 3$\sigma$ noise threshold of our observations in the Galactic centre, suggesting that the tentative detection shows signs of maser activity.  We note that our estimate is simplistic in that it does not consider the chemistry of SH and so should be treated with some caution.  

There are two reported infrared detections of SH, representing warm shocked gas in absorption. \cite{YK2000} detected the molecule in the S-type star R Andromedae with a calculated column density of 4.0$\times$10$^{20}$\,cm$^{-2}$ using the Kitt Peak 4\,m telescope.  \cite{Neufeld} detected SH using SOFIA in the diffuse cloud, W49N, and calculated the column density to be 4.6$\times$10$^{12}$\,cm$^{-2}$.    \cite{Neufeld} noted that the abundance is higher than expected, but SH is expected to have a boosted abundance in areas of warm bow shocks \citep{Neufeld,YK2000}.

In the radio, we expect to detect these low energy transitions within areas of cold gas, instead of the shocked gas in which the infrared detections were made.  Assuming standard thermal absorption mechanisms, we calculate a total column density of 2.2$\times$10$^{20}$\,cm$^{-2}$, using the partition function and upper level degeneracy, for low temperatures, as quoted in the CDMS catalogue.  This is the same order of magnitude as reported by \cite{YK2000} in the infrared, and the beam sizes are likely to be similar. 

The star 2MASS~J$17360840-2533343$, has not been part of any study and no information is known regarding its distance or velocity. However, the velocity of 26$\pm$13\,km\,s$^{-1}$ is realistic based the Galactic rotation curve in CO generated by \cite{Dame}.  There are no other known molecular transitions within 200\,km\,s$^{-1}$ of this transitions, so no other molecule can be assigned to this absorption peak.

\begin{figure*}
	\centering
       	 
	 {%
	 \includegraphics[width=1\textwidth]{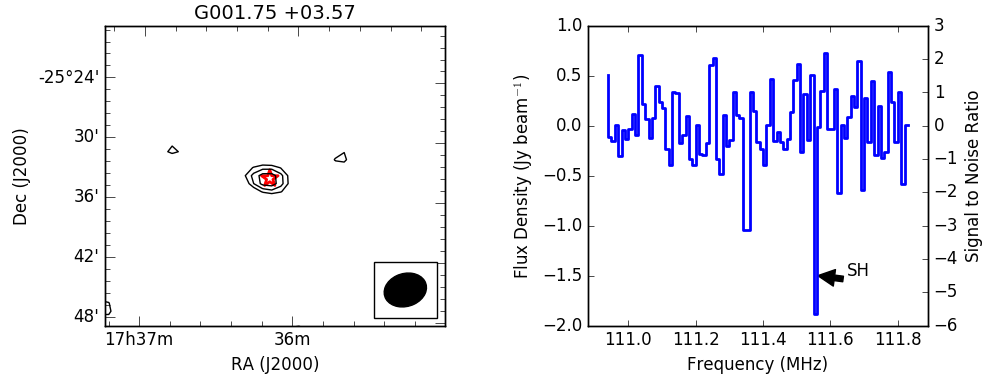}
	}
	\caption{A tentative detection of the mercapto radical (SH) with a possible association to 2MASS~J$17360840-2533343$. The image on the left is a contour plot of the detection with contours set at the 3, 4, 5, and 6\,$\sigma$ levels where 1\,$\sigma$ is 0.29\,Jy\,beam$^{-1}$. The star in the image represents the optical position of the star labeled in the upper left-hand panel and size of the star represents the astrometric uncertainty after ionospheric correction. The right-hand panel shows the spectrum from 110.9 to 111.8\,MHz at the position of 17$^{h}$36$^{m}$10$^{s}$ -25$^{\circ}$33$^{\prime}$ 51$^{\prime\prime}$(J2000).  }
\label{SH}
\end{figure*}

\begin{table*}
\small
\caption{Results for the search for the mercapto radical (SH).  The rest frequency is listed in the second row and the third row lists the upper energy level divided by the Boltzmann constant for the transition. The line strengths in terms of S$\mu^2$, with units of Debye$^2$ in the fourth row and the Einstein Coefficient (A$_{ij}$) in the fifth row. The last row lists the column density without taking into account beam dilution and assuming standard thermal emission or absorption properties (\S 3.1.2).  However, the column densities may be non-physical quantities if these represent boosted emission, such as maser activity.}

\label{SH}
%\centerline{
\begin{tabular}{lllll}
\hline
Molecule		   					&SH$^{\star}$&SH$^{\star}$&SH$^{\star}$&SH$^{\star}$ \\		
Rest Frequency (MHz)       &100.29&111.49 &111.55&122.74\\
\hline
\hline			
E$_{{\mathrm{U}}}/$ k$_{\mathrm{B}}$ (K)        &0.00                  &0.01&0.01&0.01 \\
S$\mu^{2}$ (D$^{2}$)        &0.17 &0.86 & 1.55&0.17\\
Log$_{10}$ (A$_{ij}$)         &-15.17&-14.34&$-$14.30&-15.13\\
%Obs Frequency (MHz)				&111.56	   \\
%Obs Flux Density (Jy/beam)			&$-$1.6 \\
Velocity (km s$^{-1}$)				&Not Observed&Not Detected&$-$26$\pm$13&Not Detected\\
Quantum Numbers		&$^{2}\Pi_{3/2} J=~\frac{3}{2}$, F~=1$^{+}$--~2$^{-}$&$^{2}\Pi_{3/2} J=~\frac{3}{2}$, F~=1$^{+}$--~1$^{-}$&$^{2}\Pi_{3/2}~J=\frac{3}{2}$,~F=~2~$^{+}~$--2$~^{-}$&$^{2}\Pi_{3/2}~J=\frac{3}{2}$,~F=~2~$^{+}~$--1$~^{-}$\\
%Log$_{10}$ (A$_{ij}$)				&1.55\\
%Catalogue$^{\diamondsuit}$ 			&JPL\\
Source 							&&&2MASS~J$17360840-2533343$&\\
Position (J2000)					&&&17$^{h}$36$^{m}$10$^{s}$ $-$25$^{\circ}$33$^{\prime}$ 51$^{\prime\prime}$&\\
Position (Galactic)					&&&001.75 +03.57&\\
N$_{{\mathrm{tot}}}$ (cm$^{-2}$) 		&&&2.2($\pm$0.6)$\times$10$^{20}$& \\

\hline
\end{tabular}

\small{Catalogue -- JPL -- Jet Propulsion Laboratory Spectral Line Catalogue, \citep{Pickett}\\
$^{\star}$ Molecule detected in evolved stars as per the NIST Spectral Database \citep{Lovas}}

\end{table*}

\subsubsection{Tentative Detection of Nitric Oxide (NO)}
\begin{figure*}
	\centering
       	 
	 {%
	 \includegraphics[width=0.9\textwidth]{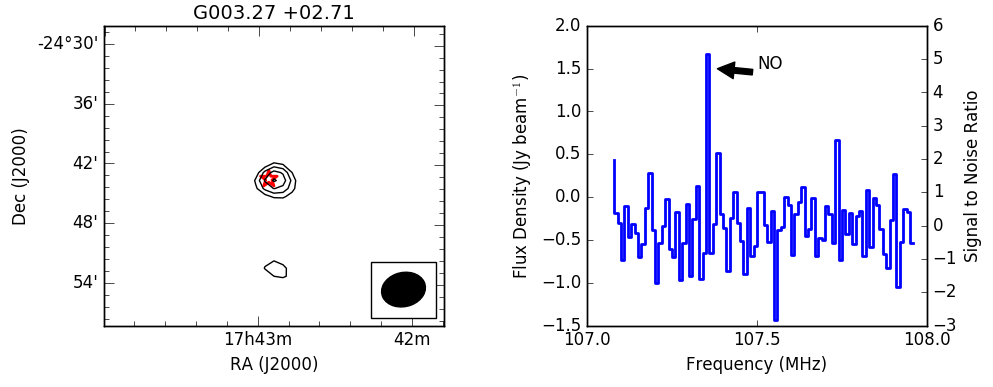}
	}
	
	\centering
	
	 {%
	 \includegraphics[width=0.9\textwidth]{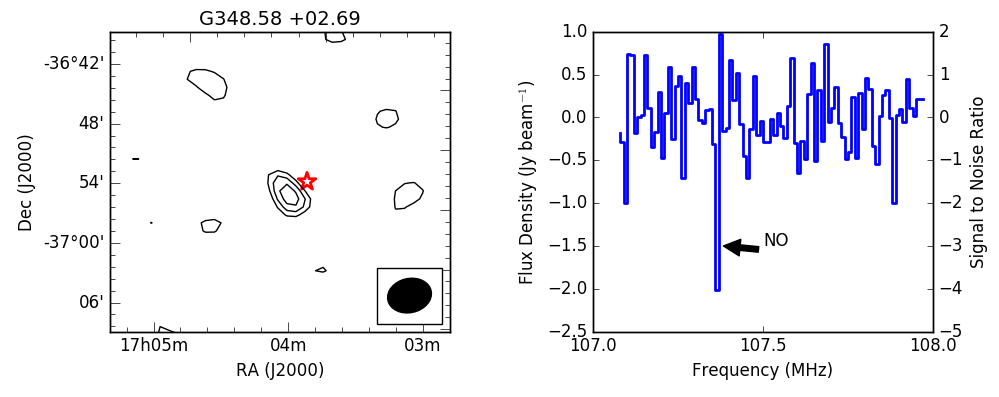}
	}
	
	\centering
	
	 {%
	 \includegraphics[width=0.9\textwidth]{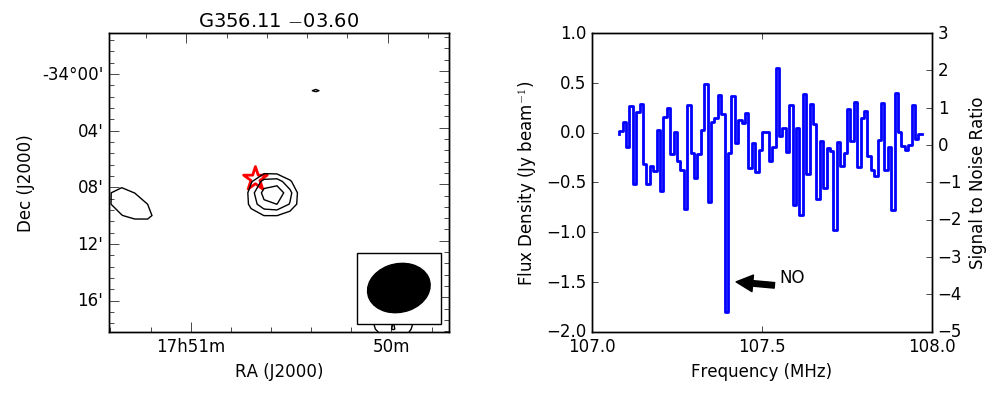}
	}
	
	\caption{Tentative detections of nitric oxide. The contours on each left-hand image are set to the 3, 4, 5, and 6\,$\sigma$ levels.  The red star in each contour plot represents the optical position of the evolved star that is a plausible association with the detection, and the size of the star represents the astrometric uncertainty after ionospheric correction.  The right-hand panel shows the full spectrum from 107\,MHz to 108\,MHz at the position of the tentative detection.  (Top) NO tentative detection at the position of 17$^{h}$42$^{m}$55$^{s}$ -24$^{\circ}$43$^{\prime}$ 9$^{\prime\prime}$ (G003.27 +02.71) with a possible association to OGLE BLG-LPV 21112. The contour map uses a 1\,$\sigma$ value of 0.38\,Jy\,beam$^{-1}$. (Middle) NO tentative detection at the position of 17$^{h}$04$^{m}$12$^{s}$ -36$^{\circ}$52$^{\prime}$ 31$^{\prime\prime}$ (G348.58 +02.69) with a possible association to TYC~$7372-165-1$. The contour map uses a 1\,$\sigma$ value of 0.50\,Jy\,beam$^{-1}$. (Bottom) NO tentative detection at the position of 17$^{h}$50$^{m}$33$^{s}$ -34$^{\circ}$07$^{\prime}$46$^{\prime\prime}$ (G356.11 $-$03.60) with a possible association to OGLE BLG-LPV 65700.  The contour map uses a 1\,$\sigma$ value of 0.40\,Jy\,beam$^{-1}$.}
	\label{NO1}
\end{figure*}

In the frequency band of these observations, there are several hyper-fine transitions of nitric oxide with an upper energy level for the transitions greater than 300\,K; so therefore we did not search at these frequencies as they are unlikely to be detected.  However, one nitric oxide transition listed in Table \ref{NO} and transitions for four nitric oxide isotopologues listed in Table \ref{Molecules} have low enough energies to be likely detectable.  The transition at 107.37\,MHz represents J=$\frac{3}{2}$, $\Omega=\frac{1}{2}$,  F=$\frac{5}{2} - \frac{5}{2}$ with an upper energy level of 7.195\,K, so is the most likely to be detected. At this frequency we have tentatively detected NO in three objects at the level of 5\,$\sigma$, as shown in Figure \ref{NO1}.

At the frequency of 107.36\,MHz, we tentatively detect emission which appears to be associated with a semi-regular pulsating star OGLE BLG-LPV 21112.  This star was identified by \cite{Soszynski13} in a survey of variable stars around the Galactic bulge.  At the frequency of 107.37 and 107.40\,MHz, we tentatively detect NO in absorption at the position of G348.58 +02.69 and G356.11 -03.60.   

Both \cite{Quintana-Lacaci} and \cite{Chen14} suggest NO is observed in the same regions as hydroxyl (OH) in shocked gas and interacts by the reaction OH~+~N~$\rightarrow$~NO~+~H.  Hydroxyl masers are found in regions of high-mass star formation, main sequence evolved stars and around the expanding circumstellar envelopes of variable red giant stars \citep{Benson90, Rudnitskij02}.  In these variable stars, \cite{Rudnitskij02} explains that a quasi-stationary layer of gas and dust, about 10 solar radii from the centre of the star, hosts a variety molecules.  Occasionally, a small shock wave of approximately 6$-$10\,km\,s$^{-1}$ crosses this layer, creating a pumping mechanism for masers to occur.   

The emission and absorption peaks tentatively detected are at 107.36\,MHz, 107.37 and 107.40\,MHz (Table \ref{NO}), could represent one of two nitric oxide transitions. The nitric oxide transition J=$\frac{17}{2}$, $\Omega=\frac{3}{2}$,  F=$\frac{19^{+}}{2} - \frac{19^{-}}{2}$ at 107.400\,MHz is a hyperfine transition with an upper energy level of 370\,K. The other transition represents J=$\frac{3}{2}$, $\Omega=\frac{1}{2}$,  F=$\frac{5}{2} - \frac{5}{2}$ at 107.368\,MHz with an upper energy level of 7.195\,K.  Since we did not detect any of the other hyperfine transitions on the individual sources and the energy of the transition at a rest frequency of 107.400\,MHz is high, we expect the emission to be associated with J=$\frac{3}{2}$, $\Omega=\frac{1}{2}$,  F=$\frac{5}{2} - \frac{5}{2}$ at 107.36\,MHz.  

When the source position for OGLE BLG-LPV 65700 is compared to the Galactic rotation curve in CO generated by \cite{Dame},  the velocity of 78$\pm$13\,km\,s$^{-1}$ is determined to be realistic. An alternative transition around 107.37\,MHz is N$^{18}$O at 107.311\,MHz, which has an upper energy of 6.86\,K and represents the J = $\frac{3}{2}$ -- $\frac{3}{2}$ , $\Omega$ = $\frac{1}{2}$, F = $\frac{5-}{2}$ -- $\frac{3+}{2}$ transition.  However, this would mean the velocity is approximately $-$221\,km\,s$^{-1}$ for each source, which is higher than standard velocities, so is an unlikely association.  

The column density for the emission and absorption lines were calculated as per Appendix 1 using the partition function and upper level degeneracy, for low temperatures, as quoted in the CDMS catalogue.   For the absorption lines, the optical depth ($\tau$) was close to unity for both objects when the continuum brightness temperature for detection location was scaled from the Haslam 408\,MHz map.  The brightness temperature for the emission line, using the Raleigh-Jeans approximation, is 4370\,K.  This suggests the emission is non-thermal and that the column density calculation may be non-representative of the environment.  %The column densities listed in Table \ref{NO}, are less than calculated by \cite{Quintana-Lacaci} in the evolved star IRC+10420, even when a similar excitation temperature is used we correct for our beam size compared to their survey. Our values are similar to other values reported in evolved stars and star forming regions \citep{Liszt78, Ziurys, Gerin}.   

\begin{table*}
\small
\caption{Results for the search for nitric oxide.  The rest frequency is listed in the second row and the third row lists the upper energy level divided by the Boltzmann constant for the transition. We quote the line strength in terms of S$\mu^{2}$ in Einstein Coefficient (A$_{ij}$) in the third and fourth row. The fifth and sixth row lists the source position in J2000 and Galactic coordinates. The last row lists the column density without taking into account beam dilution and assuming standard thermal emission or absorption properties (\S 3.1.2). However, the column densities may be non-physical quantities if these represent boosted emission, such as maser activity.}

\label{NO}
%\centerline{
\begin{tabular}{llll}
\hline
Molecule 	&	NO$^{\dagger}$${\star}$&NO$^{\dagger}$${\star}$ &NO$^{\dagger}$${\star}$\\
Rest Frequency (MHz)	&	107.37&107.37 &107.37\\
\hline
\hline
%Observed Frequency	 (MHz)	& 		& 		&	\\
%Observed Flux Density (Jy beam$^{-1}$)	&	1.9	&	-1.8	&	-2.3\\
E$_{{\mathrm{U}}}/$ k$_{\mathrm{B}}$ (K)                           &	7.25 & 	7.25 &	7.25 \\
S$\mu^{2}$ (D$^{2}$)          & 0.00 & 0.00 & 0.00\\
Log$_{10}$ (A$_{ij}$)				&	$-$18.49&$-$18.49	&$-$18.49	\\
Velocity (km s$^{-1}$)				&	$-$26$\pm$13	&	0$\pm$13	&	78$\pm$13\\
Quantum Numbers		&J=$\frac{17}{2}$, $\Omega=\frac{3}{2}$,  F=$\frac{19^{+}}{2} - \frac{19^{-}}{2}$ &J=$\frac{17}{2}$, $\Omega=\frac{3}{2}$,  F=$\frac{19^{+}}{2} - \frac{19^{-}}{2}$ &J=$\frac{17}{2}$, $\Omega=\frac{3}{2}$,  F=$\frac{19^{+}}{2} - \frac{19^{-}}{2}$ \\
%Catalogue$^{\diamondsuit}$			&	CDMS	&CDMS	&CDMS\\
Source 							&	OGLE BLG-LPV 21112	&	TYC~$7372-165-1$	&	OGLE BLG-LPV 65700\\
Position (J2000)					&	17$^{h}$42$^{m}$55$^{s}$ -24$^{\circ}$43$^{\prime}$ 9$^{\prime\prime}$		&	17$^{h}$04$^{m}$12$^{s}$ -36$^{\circ}$52$^{\prime}$ 31$^{\prime\prime}$	&	17$^{h}$50$^{m}$33$^{s}$ -34$^{\circ}$07$^{\prime}$46$^{\prime\prime}$\\
Position (Galactic)					&	003.27 +02.71	&	348.58 +02.69	&	356.11 $-$03.60\\
N$_{u}$ (cm$^{-2}$)					&	1.2($\pm$0.8)$\times$10$^{15}$			&				&				\\
N$_{{\mathrm{tot}}}$(cm$^{-2}$) 			&	1.3($\pm$0.8)$\times$10$^{14}$	&	8.5($\pm$0.8)$\times$10$^{20}$	&	1.0($\pm$0.9)$\times$10$^{21}$\\
\hline
\end{tabular}

\small{Catalogue -- CDMS -- Cologne Database for Molecular Spectroscopy \citep{Muller}\\
$^{\dagger}$ Molecule detected in star-forming regions as per the NIST Spectral Database \citep{Lovas}\\
$^{\star}$ Molecule detected in evolved stars as per the NIST Spectral Database \citep{Lovas}}

\end{table*}

\subsection{Blind Search}

A blind search in both absorption and emission was completed at a threshold of 6$\sigma$ of the local RMS measured in each final image cube.  This corresponds to a flux density limit of 0.75\,Jy\,beam$^{-1}$ in the Galactic Centre and 1.8\,Jy\,beam$^{-1}$ approximately six degrees from the Galactic Centre region.  Of the 2400 channels imaged, only 1920 channels were used in the search for spectral lines due to the aliasing effects near the edge channels (see \S 2.1 for details).  Since the spectral noise across the image is confirmed to be Gaussian, at 6$\sigma$ we can expect one false detection in the 1920 channels and 4$\times$10$^{6}$ pixels that were imaged.  %This equates to a 0.05 chance that any peak around 6$\sigma$ is noise.

Any peak over the 6$\sigma$ limit was then compared with known molecular lines from the databases CDMS, SLAIM, JPL and Top Model.  In the frequency band of 103 to 133\,MHz, there are 28 known species that have published transitions with energy in the upper excitation state less than 300\,K.  A selection of 22 of the molecules most likely to be detected based on their line strengths, energy levels and quantum mechanics are listed in Table \ref{Molecules}.  The rest frequencies for each transition were determined from a mix of laboratory experiments and theoretical modelling.  The details are found in the associated database listed in Table~\ref{Molecules}.  For any known molecular lines, the data were then correlated with objects in SIMBAD for any sources within three arc minutes of the centre pixel.

We discovered one peak above the 6\,$\sigma$ threshold at 124.91\,MHz that was associated with the known star OGLE BLG RRLYR 1344 (17$^{h}$40$^{m}$27$^{s}$ -21$^{\circ}$22$^{\prime}$ 53$^{\prime\prime}$), but was not associated with any known molecular transition.  Further laboratory experiments for molecular transitions, at these low frequencies, would be helpful to determine useful targets and integration times for more sensitive searches. 
 
To calculate the chance of a false detection corresponding to a known molecular line and known source position, we used an OH survey to approximate a source density within our field-of-view. This is a reasonable proxy because OH masers are known to occur within different astrophysical objects in which spectral lines are prominent, such as evolved stars and star-forming regions. In the OH survey by \cite{Sevenster}, a 105\,square\,degree search of the Galactic Centre yielded 286 objects hosting OH masers. If we extrapolate this to our field of 400 square degrees, we could expect around 1144 sources. Molecular lines can have a velocity of up to 150\,km\,s$^{-1}$ around the Galactic Centre \citep{Olofsson}. This velocity means that a molecular transition can shift up to 5 channels in either direction from the rest frequency.  Therefore, we would expect molecular lines to be within only 111 channels of the 1920 searched. Taking these numbers into account we would expect 0.00002 channels (or 0.0022 chance) to have a peak flux density greater than 6\,$\sigma$ and coincidentally correspond with a known molecular line and a known source.  

This survey yielded no results in emission or absorption above the 6\,$\sigma$ limit that were associated with known objects and known molecular lines.  The results in Table \ref{Molecules} represent the upper limit on the column densities for 22 molecules potentially detectable within these observations.  The limit on the column densities are calculated assuming standard thermal excitation mechanisms.  However, this may not be the case for low frequency transitions.  

\begin{table*}
\small
\caption{Molecules in the ground vibrational state that have known transitions in the 103 to 133\,MHz range with upper level energies less than 300\,K.  The first column lists the name of the molecule.  We also note molecules detected at other frequencies in star-forming regions ($\dagger$) and evolved stars ($\star$) in this column. The rest frequency is listed in the second column and the third column lists the upper energy level divided by the Boltzmann constant for the transition. We quote the line strengths in terms of S$\mu^2$, with units of Debye$^2$ in the fourth column and the Einstein Coefficient (A$_{ij}$) in the fifth column. The sixth column lists the catalogue used to source the data and the last column lists the upper limit on the column density (\S 3.1.2) for the most probable line of each molecule. Note that the detectability of any spectral line will depend on the S$\mu^{2}$ and A$_{ij}$ value and the relative abundance of the molecule. However, relative abundances for most of the molecules listed are not known.}

\label{Molecules}
%\centerline{
\begin{tabular}{lccccccll}
\hline
\hline
Molecule								&	Rest Frequency 	&E$_{{\mathrm{U}}}/$ k$_{\mathrm{B}}$ (K)	&S$\mu^{2}$& 	Log$_{10}$ (A$_{ij}$) &	Catalogue$^{\diamondsuit}$ & 	N$_{u}$		&N$_{{\mathrm{tot}}}$ \\
									&	(MHz)		&(K)		&(D$^{2}$)    &					  &						 &(cm$^{-2}$)		&	(cm$^{-2}$)\\
\hline
CH$_{2}$OO$^{\dagger}$					&	104.52	&		152.20	&		58.96 	&-13.54		& SLAIM	&		Flagged				&Flagged\\
CH$_{3}$O$^{13}$CHO$^{\dagger}$			&	113.86	&		4.38&		 	0.10		&-15.45		&Top Model		&$<$4.24$\times$10$^{12}$			&$<$1.64$\times$10$^{13}$\\
l-C$_{3}$H$^{\dagger\star}$				&	107.98	&		32.93	&		9.20		&-13.91		&JPL			&Flagged				& Flagged\\
l-C$_{4}$H$_{2}$ 						&      107.33	& 		144.09 	& 		7.76 		&-14.57		& SLAIM 			&$<$8.59$\times$10$^{12}$			& $<$1.05$\times$10$^{17}$\\
                              						& 	129.91  	& 		153.09 	& 		7.39 		&-14.36		& SLAIM 			&Flagged				& Flagged\\
t-DCOOH$^{\dagger}$					&	124.80	&		42.40	&		5.41		&-14.03		&JPL			&$<$6.10$\times$10$^{12}$			&$<$3.48$\times$10$^{13}$\\
cis-H$^{13}$COOH$^{\dagger}$			&	105.78	&		20.38	& 		16.37  	&-13.49		&JPL			&Flagged				&Flagged\\
									&	110.82	&		182.72	&		14.42 	&-14.13		&JPL			&Flagged				&Flagged\\
cis-HCOOD$^{\dagger}$					&	106.53	&		190.60	&		13.55	&-14.24		& JPL							& \\
									&	118.78	&		76.16	&		13.31	&-13.86		&JPL			&$<$9.39$\times$10$^{11}$			& $<$6.64$\times$10$^{13}$ \\
H$_{2}$$^{13}$CO$^{\dagger}$			&	116.90	&		201.85	&		39.28	&-12.90		&CDMS			&$<$2.44$\times$10$^{12}$			& $<$1.72$\times$10$^{19}$\\
l-H$_{2}$CCCO						&	111.69	&		56.58	&		14.56 	&-14.62		&JPL			&$<$2.28$\times$10$^{10}$			& $<$1.86$\times$10$^{11}$\\
									&	128.93	&		7.57		&		20.92 	&-13.24		&JPL			&Flagged				& Flagged\\
H$_{2}$CS$^{\dagger\star}$				&	104.46	&		77.39	&		3.99 		&-13.87		&CDMS			&$<$5.52$\times$10$^{12}$	& $<$5.39$\times$10$^{14}$\\         
H$_{2}$C$_{2}$S						&	114.96	&		14.03	&		4.59		&-14.04		&CDMS			&$<$1.32$\times$10$^{13}$	&$<$5.32$\times$10$^{13}$\\
HCCCH$_{2}$OH						&	118.06	&		195.95	&		1.67		&-15.23		&JPL							& \\
									&	125.00	&		71.74	&		1.72	 	&-14.77		& JPL			&$<$7.47$\times$10$^{12}$			& $<$1.31$\times$10$^{14}$\\
									&	125.91	&		134.32	&		1.87		&-14.95		& JPL& 			Flagged				&Flagged\\
									&	129.90	&		239.05	&		2.18		&-15.01		& JPL			&Flagged				& Flagged\\
HCCCHO  							&  	107.54 	&		256.84  	& 		5.80 		&-14.86		& SLAIM 							& \\
									&  	117.22 	& 		95.85 	& 		5.71 		&-14.51		& SLAIM 			&$<$1.71$\times$10$^{12}$			& $<$1.71$\times$10$^{15}$\\
t-HONO								&	105.03	&		22.35	&		4.43		&-14.07		&JPL			&Flagged				& Flagged \\
t-HCOOD$^{\dagger}$					&	106.19	&		123.92	&		4.95		&-14.56		&JPL			&$<$6.16$\times$10$^{12}$			& $<$1.12$\times$10$^{14}$\\
HDCO$^{\dagger}$						&	129.09	&		41.13	&		18.23	&-13.04		&JPL			&$<$2.21$\times$10$^{12}$			& $<$4.00$\times$10$^{12}$\\
NH$_{2}$CH$_{2}$CH$_{2}$OH			&	103.09	&		124.82	&		31.85	&-14.00		&JPL			&Flagged				& Flagged\\
									
									&	106.87	&		82.74	&		29.42	&-13.90		&JPL			&Flagged				& Flagged\\
									&	107.84	&		100.59	&		5.33		&-14.72		&JPL& 							\\
									&	114.75	&		41.80	&		4.12		&-14.57		&JPL& 			$<$1.69$\times$10$^{12}$			&$<$1.37$\times$10$^{13}$\\
									&	123.20	&		171.85	&		6.17		&-14.60		&JPL& 							\\

$^{15}$NO							&	105.74	&		18.61	&		$<$0.01	&-18.31		&  CDMS			&Flagged				& Flagged\\
									&	120.27	&		6.99		&		$<$0.01 	&-18.93		&CDMS 			&$<$5.11$\times$10$^{12}$			& $<$7.12$\times$10$^{13}$\\
N$^{17}$O							&	105.34	&		0.01		&		0.01		&-16.48		&CDMS			&$<$5.84$\times$10$^{14}$			& $<$2.07$\times$10$^{14}$\\	
									&	108.52	&		191.69	&		0.01	 	&-16.53		&CDMS							& \\
									&	109.20	&		0.01		&		0.04 		&-16.23		&CDMS							& \\
									&	114.46	&		191.69	&		0.01		&-16.39		&CDMS			&Flagged				& Flagged\\

N$^{18}$O							&	107.31	&		6.87		&		$<$0.01 	&-18.49		&CDMS			&$<$5.73$\times$10$^{12}$			& $<$5.99$\times$10$^{13}$ \\
$^{15}$N$^{17}$O						&	109.17	&		54.93	&		$<$0.01 	&-17.13		&	CDMS& 						\\
									&	109.34	&		54.21	&		$<$0.01	&-17.14		&	CDMS		&$<$5.63$\times$10$^{14}$			& $<$2.78$\times$10$^{16}$\\
									&	109.48	&		142.24	&		0.01		&-17.12		&CDMS			&Flagged				& Flagged\\
									&	129.78	&		79.04	&		0.01	 	&-16.97		&CDMS							& \\
c-SiC$_{3}$$^{\star}$					&	112.70	&		35.16	&		48.77	&-13.32		& SLAIM 			& $<$7.79$\times$10$^{11}$			& $<$7.54$\times$10$^{12}$	\\
									&	114.62	&		86.28	&		50.42	&-13.52		& SLAIM 			&Flagged				&Flagged\\
									&	124.02	&		292.73	&		56.67	&-13.66		& SLAIM			&Flagged				&Flagged\\

\hline
\end{tabular}

\small{$\diamondsuit$SLAIM-Spectral Line Atlas of Interstellar Molecules (NRAO Splatalogue), JPL-Jet Propulsion Laboratory Spectral Line Catalogue, \citep{Pickett}, CDMS-Cologne Database for Molecular Spectroscopy \citep{Muller}, TopModel-\citep{Carvajal}\\
$^{\dagger}$ Molecule detected in star-forming regions as per the NIST Spectral Database \citep{Lovas}\\
$^{\star}$ Molecule detected in evolved stars as per the NIST Spectral Database \citep{Lovas}}

\end{table*}

%\section{Discussion \& Summary}

%%%%%%%%%%%%%%%%%%%%%%%%%%%%%%%%%%%%%%%%%%%%%%%%%%%%%%%%%%%%%%%%%%%%%%%%%%%%%%%%%
%%%									CONCLUSTION																		%%
%%%%%%%%%%%%%%%%%%%%%%%%%%%%%%%%%%%%%%%%%%%%%%%%%%%%%%%%%%%%%%%%%%%%%%%%%%%%%%%%%

\section{Discussion}
This survey is the first reported molecular search at low frequency and the goal was to assess the feasibility of using the MWA for this style of work. Within the 400\,square\,degree field-of-view, we would expect up to 1,100 sources (as discussed in \S 4.2) to contain molecular rich environments in which molecular transitions could be detected.  The telescope's wide field-of-view, along with the extreme radio quiet of the Murchison Radio-astronomy Observatory \citep{OffriingaRFI}, makes it a great survey instrument to find new and interesting molecular regions. The search yielded tentative detections of nitric oxide in three objects at 107.37\,MHz and the mercapto radical at 111.55\,MHz in one evolved star. Here we attempt to explain the difference between the number of objects with tentative detections and the total potential sources within the field of view. 

At radio frequencies, the ratio of photon energy to kinetic energy ($h\nu/kT$) at a temperature T is very small ($<<$1).  This makes almost every object a thermal radio source, as the brightness of the blackbody emitter is proportional to the square of the frequency ($\nu^{2}$). For radio spectral lines, $h\nu/kT$ $<<1$ lowers the opacity limits and makes the emission strength independent of temperature of the emitting gas. The emission is then proportional to the number of atoms in the proper stage of ionisation \citep{CondonERA} and allows for maser emission with only a small population inversion \citep{SpitzerISM}. Therefore, at frequencies less than 1\,GHz, we are more sensitive to stimulated emission and maser transitions.   

The set-up of the MWA used during this survey offers a few challenges.  Only 78\% of the bandpass was imaged due to aliasing from the polyphase filter bank.  This reduced the number of potentially detectable known transitions by 36\%.  Also, with the 3\,km baselines, the FWHM of the synthesised beam is 3\,arc\,min, resulting in a large beam dilution factor for objects in or near the Galactic centre, as determined by a ratio of solid angles. 

\begin{figure}
	 \includegraphics[width=0.48\textwidth]{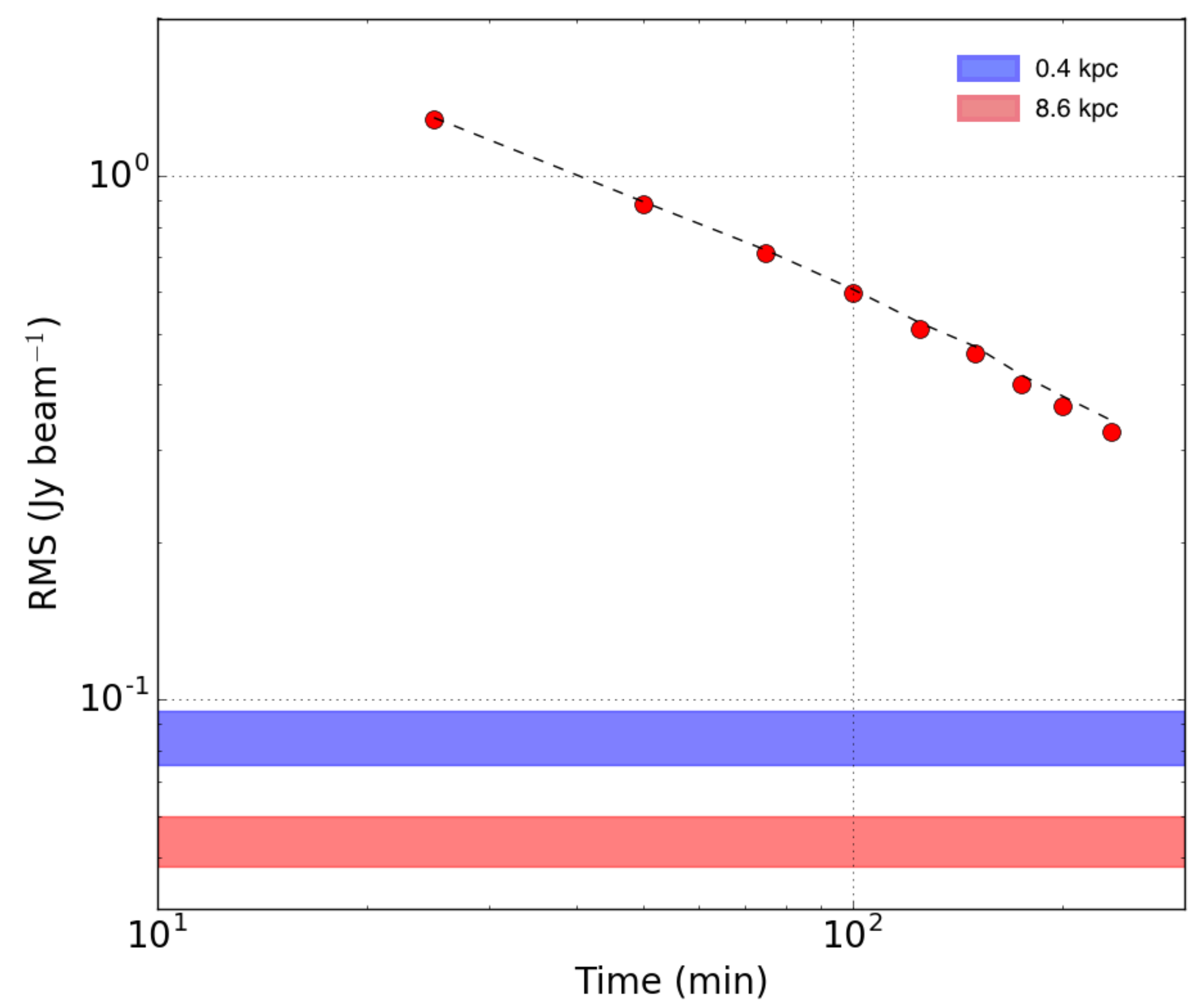}
	\caption{A plot of RMS noise against integration time, where red dots show the RMS versus time and the dashed line indicates the trend of square root of time. Therefore, more time integration on the same source should give improvement for possible detections. The blue and red shaded bars show the estimated sensitivity to start detecting thermal emission at a distance of 0.4\,kpc and 8\,kpc (approximate distance to the Galactic Plane) respectively. }
	\label{Noise}
\end{figure}

The beam dilution could be mitigated by increased integration time.  For this survey, the addition of the data observed in 2016 and integrated with the observations from 2014, demonstrated the continuation of the sensitivity increase as a function of square root of time, as expected from the radiometer equation.  In order to obtain the sensitivity to match the column densities set by higher frequency observations, we would need to observe for 65 to 100\,hours for a five sigma detection (see Figure \ref{Noise}).  This is the same order of magnitude determined by \cite{Codella14} for integration time required to observe complex organic molecules with SKA1-Mid at around 1\,GHz.  However, any stimulated emission would boost the detectability and therefore, reduce the required amount of integration time.  In the context of this survey, with four hours of total integration time, we are only sensitive to unique stimulated environments.

The MWA is undergoing an upgrade to add baselines of lengths up to 6\,km, which reduces the beam size and therefore the beam dilution.  This would increase the angular resolution to 1.6\,arc\,minutes, could increase the sensitivity to less abundant molecules and increase our ability to resolve the regions in which the molecules are detected.  

Telescopes observing around 100\,MHz are strongly influenced by the ionosphere, causing changes in the apparent position of the sources. As discussed in \S 2.1.2, this was corrected with a single correction factor for each snapshot observation. However, with longer baselines, not all of the dipoles will be observing the same isoplanatic patch of the ionosphere and different tools will need to be utilised.

The biggest challenge in this style of work with the MWA and possibly with the future SKA-Low is the spectral sensitivity.  The aliasing from the polyphase filterbank reduced the number of transitions we could detect and the spectral shape of each fine channel reduced our sensitivity to transitions near the edges of the fine channels.  This is especially apparent with the channel resolution of 26\,km\,s$^{-1}$, where most spectral line transitions have a velocity of only a few kilometres per second. 

Overall, the setup of the MWA during this survey, combined with the amount of integration time and beam dilution, made us sensitive only to bright stimulated transitions with broad velocity components.  Although we have estimated there are likely upwards of 1100 potential sources with potential molecular transitions, these restrictions reduce our chance of detection, but it is difficult to predict by how much, as the expected intensity of the transitions and abundance of the molecules is unknown.

\section{Summary \& Conclusions} \label{Conclusion}

The goal of this pilot survey was to assess the feasibility of using low frequency telescopes to observe spectral lines, develop a method for observing molecular lines with the MWA and search for molecular lines in and around the Galactic Centre.  From the setup of the MWA described in \S 2, we are only sensitive to bright molecular transitions with broad excitation.  Therefore, there is likely a selection effect to only detect molecules in extraordinary environments.  The excitation in these environments is likely non-thermal in emission, as the brightness temperatures are determined to be $>$1000\,K, suggesting either quasi-thermal or maser type emission and absorption.  

 In this survey, any peak of over 6\,$\sigma$ local RMS in the blind search and 5\,$\sigma$ in the targeted search for NO and SH, was cross-matched with known molecular lines and known sources.  We have tentatively detected SH and NO in evolved stars at a level of $5\sigma$.   However, since both molecules only have a single transition observed in a single channel, they are considered tentative.  The column densities calculated are similar to those reported in other surveys, when a correction for beam size is applied. 

The RMS decreases with respect to the square root of time. To match the column densities reported in other radio surveys for NO, and to take into account our beam dilution, we would need approximately 30~hours of observations with the MWA to obtain a five sigma detection, assuming standard thermal emission or absorption processes at distances of around 400\,pc and 65--100\,hours for detections at distances closer to the Galactic Centre.  Therefore, the next stage of assessing the low-frequency sky is to complete a full survey of the Orion region from 99--270\,MHz.  Orion is known as a unique environment with a mix of star-forming and evolved stars, at a distance of approximately 400pc. 

At low radio frequencies we are more likely to detect maser activity than is possible at the higher frequency as the emission strengths are independent of temperature of the emitting gas and maser activity is possible with only a small population inversion \citep{SpitzerISM}.  This significantly increases the detectability of molecular transitions.  However, it is unknown if all detectable transitions would be stimulated or not.

Overall, the analysis and tentative detections show that it is feasible to detect molecules with the new generation of low-frequency instruments.

\section*{Acknowledgments}
CDT would like to thank Steven Tremblay for very helpful discussions about low frequency radio astronomy.  Parts of this research were conducted by the Australian Research Council Centre of Excellence for All-sky Astrophysics (CAASTRO), through project number CE110001020. This work was supported by resources provided by the Pawsey Supercomputing Centre with funding from the Australian Government and the Government of Western Australia.

This scientific work makes use of the Murchison Radio-astronomy Observatory, operated by CSIRO. We acknowledge the Wajarri Yamatji people as the traditional owners of the Observatory site. Support for the operation of the MWA is provided by the Australian Government (NCRIS), under a contract to Curtin University administered by Astronomy Australia Limited. 

%%%%%%%%%%%%%%%%%%%%%%%%%%%%%%%%%%%%%%%%%%%%%%%%%%



%\bibliographystyle{mnras}

\bibliographystyle{mnras.bst}
\bibliography{research3}

\begin{thebibliography}{}
\makeatletter
\relax
\def\mn@urlcharsother{\let\do\@makeother \do\$\do\&\do\#\do\^\do\_\do\%\do\~}
\def\mn@doi{\begingroup\mn@urlcharsother \@ifnextchar [ {\mn@doi@}
  {\mn@doi@[]}}
\def\mn@doi@[#1]#2{\def\@tempa{#1}\ifx\@tempa\@empty \href
  {http://dx.doi.org/#2} {doi:#2}\else \href {http://dx.doi.org/#2} {#1}\fi
  \endgroup}
\def\mn@eprint#1#2{\mn@eprint@#1:#2::\@nil}
\def\mn@eprint@arXiv#1{\href {http://arxiv.org/abs/#1} {{\tt arXiv:#1}}}
\def\mn@eprint@dblp#1{\href {http://dblp.uni-trier.de/rec/bibtex/#1.xml}
  {dblp:#1}}
\def\mn@eprint@#1:#2:#3:#4\@nil{\def\@tempa {#1}\def\@tempb {#2}\def\@tempc
  {#3}\ifx \@tempc \@empty \let \@tempc \@tempb \let \@tempb \@tempa \fi \ifx
  \@tempb \@empty \def\@tempb {arXiv}\fi \@ifundefined
  {mn@eprint@\@tempb}{\@tempb:\@tempc}{\expandafter \expandafter \csname
  mn@eprint@\@tempb\endcsname \expandafter{\@tempc}}}

\bibitem[\protect\citeauthoryear{{Benson}, {Little-Marenin}, {Woods},
  {Attridge}, {Blais}, {Rudolph}, {Rubiera}  \& {Keefe}}{{Benson}
  et~al.}{1990}]{Benson90}
{Benson} P.~J.,  {Little-Marenin} I.~R.,  {Woods} T.~C.,  {Attridge} J.~M.,
  {Blais} K.~A.,  {Rudolph} D.~B.,  {Rubiera} M.~E.,   {Keefe} H.~L.,  1990,
  \mn@doi [SPJS] {10.1086/191526}, \href
  {http://adsabs.harvard.edu/abs/1990ApJS...74..911B} {74, 911}

\bibitem[\protect\citeauthoryear{{Bowman}, {Cairns}, {Kaplan}  et~al.}{{Bowman}
  et~al.}{2013}]{Bowman}
{Bowman} J.~D.,  {Cairns} I.,  {Kaplan} D.~L.,   et~al., 2013, \mn@doi [PASA]
  {10.1017/pas.2013.009}, \href
  {http://adsabs.harvard.edu/abs/2013PASA...30...31B} {30, e031}

\bibitem[\protect\citeauthoryear{{Briggs}}{{Briggs}}{1995}]{Briggs}
{Briggs} D.~S.,  1995, in American Astronomical Society Meeting Abstracts.
  p.~1444

\bibitem[\protect\citeauthoryear{{Bruderer}, {Benz}, {St{\"a}uber}  \&
  {Doty}}{{Bruderer} et~al.}{2010}]{Bruderer}
{Bruderer} S.,  {Benz} A.~O.,  {St{\"a}uber} P.,   {Doty} S.~D.,  2010, \mn@doi
  [APJ] {10.1088/0004-637X/720/2/1432}, \href
  {http://adsabs.harvard.edu/abs/2010ApJ...720.1432B} {720, 1432}

\bibitem[\protect\citeauthoryear{{Carvajal}, {Kleiner}  \&
  {Demaison}}{{Carvajal} et~al.}{2010}]{Carvajal}
{Carvajal} M.,  {Kleiner} I.,   {Demaison} J.,  2010, \mn@doi [APJS]
  {10.1088/0067-0049/190/2/315}, \href
  {http://adsabs.harvard.edu/abs/2010ApJS..190..315C} {190, 315}

\bibitem[\protect\citeauthoryear{{Chambers}, {Yusef-Zadeh}  \&
  {Ott}}{{Chambers} et~al.}{2014}]{Chambers}
{Chambers} E.~T.,  {Yusef-Zadeh} F.,   {Ott} J.,  2014, \mn@doi [AAP]
  {10.1051/0004-6361/201322752}, \href
  {http://adsabs.harvard.edu/abs/2014A%26A...563A..68C} {563, A68}

\bibitem[\protect\citeauthoryear{{Chen} et~al.,}{{Chen} et~al.}{2014}]{Chen14}
{Chen} J.-H.,  et~al., 2014, \mn@doi [APJ] {10.1088/0004-637X/793/2/111}, \href
  {http://adsabs.harvard.edu/abs/2014ApJ...793..111C} {793, 111}

\bibitem[\protect\citeauthoryear{{Codella}, {Podio}, {Fontani}
  et~al.}{{Codella} et~al.}{2015}]{Codella14}
{Codella} C.,  {Podio} L.,  {Fontani} F.,   et~al., 2015, Advancing
  Astrophysics with the Square Kilometre Array (AASKA14), \href
  {http://adsabs.harvard.edu/abs/2015aska.confE.123C} {p.~123}

\bibitem[\protect\citeauthoryear{{Comito}, {Schilke}, {Gerin}, {Phillips},
  {Zmuidzinas}  et~al.}{{Comito} et~al.}{2003}]{Comito}
{Comito} C.,  {Schilke} P.,  {Gerin} M.,  {Phillips} T.~G.,  {Zmuidzinas} J.,
  et~al., 2003, \mn@doi [AAP] {10.1051/0004-6361:20030293}, \href
  {http://adsabs.harvard.edu/abs/2003A%26A...402..635C} {402, 635}

\bibitem[\protect\citeauthoryear{{Condon} \& {Ransom}}{{Condon} \&
  {Ransom}}{2016}]{CondonERA}
{Condon} J.~J.,  {Ransom} S.~M.,  2016, {Essential Radio Astronomy}

\bibitem[\protect\citeauthoryear{{Cornwell}, {Golap}  \&
  {Bhatnagar}}{{Cornwell} et~al.}{2005}]{Cornwell05}
{Cornwell} T.~J.,  {Golap} K.,   {Bhatnagar} S.,  2005, \mn@doi [Astrophysics
  and Space Science Proceedings] {10.1109/ICASSP.2005.1416440}, \href
  {http://adsabs.harvard.edu/abs/2005assp.conf..861C} {5}

\bibitem[\protect\citeauthoryear{{Cosmovici}}{{Cosmovici}}{1979}]{Cosmovici79}
{Cosmovici} C.~B.,  1979, in {McCarthy} M.~F.,  {Philip} A.~G.~D.,   {Coyne}
  G.~V.,  eds,  Ricerche Astronomiche Vol. 9, IAU Colloq. 47: Spectral
  Classification of the Future. pp 439--456

\bibitem[\protect\citeauthoryear{{Cunningham} et~al.,}{{Cunningham}
  et~al.}{2007}]{Cunningham07}
{Cunningham} M.~R.,  et~al., 2007, \mn@doi [MNRAS]
  {10.1111/j.1365-2966.2007.11504.x}, \href
  {http://adsabs.harvard.edu/abs/2007MNRAS.376.1201C} {376, 1201}

\bibitem[\protect\citeauthoryear{{Dame}, {Hartmann}  \& {Thaddeus}}{{Dame}
  et~al.}{2001}]{Dame}
{Dame} T.~M.,  {Hartmann} D.,   {Thaddeus} P.,  2001, \mn@doi [APJ]
  {10.1086/318388}, \href {http://adsabs.harvard.edu/abs/2001ApJ...547..792D}
  {547, 792}

\bibitem[\protect\citeauthoryear{{Danilovich}, {De Beck}, {Black}, {Olofsson}
  \& {Justtanont}}{{Danilovich} et~al.}{2016}]{Danilovich}
{Danilovich} T.,  {De Beck} E.,  {Black} J.~H.,  {Olofsson} H.,   {Justtanont}
  K.,  2016, \mn@doi [AAP] {10.1051/0004-6361/201527943}, \href
  {http://adsabs.harvard.edu/abs/2016A%26A...588A.119D} {588, A119}

\bibitem[\protect\citeauthoryear{{Dawson}, {Walsh}, {Jones}  et~al.}{{Dawson}
  et~al.}{2014}]{Dawson}
{Dawson} J.~R.,  {Walsh} A.~J.,  {Jones} P.~A.,   et~al., 2014, \mn@doi [MNRAS]
  {10.1093/mnras/stu032}, \href
  {http://adsabs.harvard.edu/abs/2014MNRAS.439.1596D} {439, 1596}

\bibitem[\protect\citeauthoryear{{Gerin}, {Viala}, {Pauzat}  \&
  {Ellinger}}{{Gerin} et~al.}{1992}]{Gerin}
{Gerin} M.,  {Viala} Y.,  {Pauzat} F.,   {Ellinger} Y.,  1992, AAP, \href
  {http://adsabs.harvard.edu/abs/1992A%26A...266..463G} {266, 463}

\bibitem[\protect\citeauthoryear{{Godard}, {Falgarone}, {Gerin}
  et~al.}{{Godard} et~al.}{2012}]{Godard12}
{Godard} B.,  {Falgarone} E.,  {Gerin} M.,   et~al., 2012, \mn@doi [AAP]
  {10.1051/0004-6361/201117664}, \href
  {http://adsabs.harvard.edu/abs/2012A%26A...540A..87G} {540, A87}

\bibitem[\protect\citeauthoryear{{Gorai}, {Das}, {Das}, {Sivaraman}, {Etim}  \&
  {Chakrabarti}}{{Gorai} et~al.}{2016}]{Gorai}
{Gorai} P.,  {Das} A.,  {Das} A.,  {Sivaraman} B.,  {Etim} E.~E.,
  {Chakrabarti} S.~K.,  2016, preprint, \href
  {http://adsabs.harvard.edu/abs/2016arXiv161202688G} {} (\mn@eprint {arXiv}
  {1612.02688})

\bibitem[\protect\citeauthoryear{{Gupta}}{{Gupta}}{2014}]{uGMRT}
{Gupta} Y.,  2014, in Astronomical Society of India Conference Series. pp
  441--447

\bibitem[\protect\citeauthoryear{{Hancock}, {Murphy}, {Gaensler}, {Hopkins}  \&
  {Curran}}{{Hancock} et~al.}{2012}]{Hancock12}
{Hancock} P.~J.,  {Murphy} T.,  {Gaensler} B.~M.,  {Hopkins} A.,   {Curran}
  J.~R.,  2012, \mn@doi [MNRAS] {10.1111/j.1365-2966.2012.20768.x}, \href
  {http://adsabs.harvard.edu/abs/2012MNRAS.422.1812H} {422, 1812}

\bibitem[\protect\citeauthoryear{{Herbst} \& {van Dishoeck}}{{Herbst} \& {van
  Dishoeck}}{2009}]{Herbst09}
{Herbst} E.,  {van Dishoeck} E.~F.,  2009, \mn@doi [ARAA]
  {10.1146/annurev-astro-082708-101654}, \href
  {http://adsabs.harvard.edu/abs/2009ARA%26A..47..427H} {47, 427}

\bibitem[\protect\citeauthoryear{{Hildebrand}}{{Hildebrand}}{1983}]{Hildebrand}
{Hildebrand} R.~H.,  1983, QJRAS, \href
  {http://adsabs.harvard.edu/abs/1983QJRAS..24..267H} {24, 267}

\bibitem[\protect\citeauthoryear{{Hindson}, {Johnston-Hollitt},
  {Hurley-Walker}, {Callingham}, {Su}  et~al.}{{Hindson}
  et~al.}{2016}]{HindsonL}
{Hindson} L.,  {Johnston-Hollitt} M.,  {Hurley-Walker} N.,  {Callingham} J.~R.,
   {Su} H.,   et~al., 2016, \mn@doi [PASA] {10.1017/pasa.2016.19}, \href
  {http://adsabs.harvard.edu/abs/2016PASA...33...20H} {33, e020}

\bibitem[\protect\citeauthoryear{{Hurley-Walker}, {Morgan}, {Wayth}
  et~al.}{{Hurley-Walker} et~al.}{2014}]{HW14}
{Hurley-Walker} N.,  {Morgan} J.,  {Wayth} R.~B.,   et~al., 2014, \mn@doi
  [PASA] {10.1017/pasa.2014.40}, \href
  {http://adsabs.harvard.edu/abs/2014PASA...31...45H} {31, e045}

\bibitem[\protect\citeauthoryear{{Hurley-Walker}, {Callingham}, {Hancock},
  {Franzen}, {Hindson}  et~al.}{{Hurley-Walker} et~al.}{2017}]{GLEAM}
{Hurley-Walker} N.,  {Callingham} J.~R.,  {Hancock} P.~J.,  {Franzen} T.~M.~O.,
   {Hindson} L.,   et~al., 2017, \mn@doi [MNRAS] {10.1093/mnras/stw2337}, \href
  {http://adsabs.harvard.edu/abs/2017MNRAS.464.1146H} {464, 1146}

\bibitem[\protect\citeauthoryear{{Isella}, {Hull}, {Moullet}  et~al.}{{Isella}
  et~al.}{2015}]{Isella15}
{Isella} A.,  {Hull} C.~L.~H.,  {Moullet} A.,   et~al., 2015, preprint, \href
  {http://adsabs.harvard.edu/abs/2015arXiv151006444I} {} (\mn@eprint {arXiv}
  {1510.06444})

\bibitem[\protect\citeauthoryear{{Jackson}, {Rathborne}, {Foster}, {Whitaker},
  {Sanhueza}  \& {et al}}{{Jackson} et~al.}{2013}]{Jackson}
{Jackson} J.~M.,  {Rathborne} J.~M.,  {Foster} J.~B.,  {Whitaker} J.~S.,
  {Sanhueza} P.,   {et al} 2013, \mn@doi [PASA] {10.1017/pasa.2013.37}, \href
  {http://adsabs.harvard.edu/abs/2013PASA...30...57J} {30, 57}

\bibitem[\protect\citeauthoryear{{Jones}, {Burton}, {Cunningham},
  {Requena-Torres}, {Menten}  et~al.}{{Jones} et~al.}{2012}]{Jones12}
{Jones} P.~A.,  {Burton} M.~G.,  {Cunningham} M.~R.,  {Requena-Torres} M.~A.,
  {Menten} K.~M.,   et~al., 2012, \mn@doi [MNRAS]
  {10.1111/j.1365-2966.2011.19941.x}, \href
  {http://adsabs.harvard.edu/abs/2012MNRAS.419.2961J} {419, 2961}

\bibitem[\protect\citeauthoryear{{Jones}, {Burton}, {Cunningham}, {Tothill}  \&
  {Walsh}}{{Jones} et~al.}{2013}]{Jones13}
{Jones} P.~A.,  {Burton} M.~G.,  {Cunningham} M.~R.,  {Tothill} N.~F.~H.,
  {Walsh} A.~J.,  2013, \mn@doi [MNRAS] {10.1093/mnras/stt717}, \href
  {http://adsabs.harvard.edu/abs/2013MNRAS.433..221J} {433, 221}

\bibitem[\protect\citeauthoryear{{Jordan}, {Walsh}, {Lowe}  et~al.}{{Jordan}
  et~al.}{2015}]{Jordan15}
{Jordan} C.~H.,  {Walsh} A.~J.,  {Lowe} V.,   et~al., 2015, \mn@doi [MNRAS]
  {10.1093/mnras/stv178}, \href
  {http://adsabs.harvard.edu/abs/2015MNRAS.448.2344J} {448, 2344}

\bibitem[\protect\citeauthoryear{{Large}, {Mills}, {Little}, {Crawford}  \&
  {Sutton}}{{Large} et~al.}{1981}]{Large}
{Large} M.~I.,  {Mills} B.~Y.,  {Little} A.~G.,  {Crawford} D.~F.,   {Sutton}
  J.~M.,  1981, \mn@doi [MNRAS] {10.1093/mnras/194.3.693}, \href
  {http://adsabs.harvard.edu/abs/1981MNRAS.194..693L} {194, 693}

\bibitem[\protect\citeauthoryear{{Lonsdale}, {Cappallo}, {Morales}
  et~al.}{{Lonsdale} et~al.}{2009}]{Lonsdale}
{Lonsdale} C.~J.,  {Cappallo} R.~J.,  {Morales} M.~F.,   et~al., 2009, \mn@doi
  [IEEE Proceedings] {10.1109/JPROC.2009.2017564}, \href
  {http://adsabs.harvard.edu/abs/2009IEEEP..97.1497L} {97, 1497}

\bibitem[\protect\citeauthoryear{{Lovas}, {Coursey}, {Kotochigova}, {Chang},
  {Olsen}  \& {Dragoset}}{{Lovas} et~al.}{2003}]{Lovas}
{Lovas} F.~J.,  {Coursey} J.,  {Kotochigova} S.,  {Chang} J.,  {Olsen} K.,
  {Dragoset} R.,  2003, Triatomic Spectral Database (version 2.0), \url
  {http://physics.nist.gov/cgi-bin/micro/table5/start.pl}

\bibitem[\protect\citeauthoryear{{Mangum} \& {Shirley}}{{Mangum} \&
  {Shirley}}{2015}]{Mangum}
{Mangum} J.~G.,  {Shirley} Y.~L.,  2015, \mn@doi [PASP] {10.1086/680323}, \href
  {http://adsabs.harvard.edu/abs/2015PASP..127..266M} {127, 266}

\bibitem[\protect\citeauthoryear{{McGuire}, {Carroll}, {Loomis}, {Finneran},
  {Jewell}, {Remijan}  \& {Blake}}{{McGuire}
  et~al.}{2016}]{2016Sci...352.1449M}
{McGuire} B.~A.,  {Carroll} P.~B.,  {Loomis} R.~A.,  {Finneran} I.~A.,
  {Jewell} P.~R.,  {Remijan} A.~J.,   {Blake} G.~A.,  2016, \mn@doi [Science]
  {10.1126/science.aae0328}, \href
  {http://adsabs.harvard.edu/abs/2016Sci...352.1449M} {352, 1449}

\bibitem[\protect\citeauthoryear{{Menten}}{{Menten}}{2004}]{Menten}
{Menten} K.~M.,  2004, in {Pfalzner} S.,  {Kramer} C.,  {Staubmeier} C.,
  {Heithausen} A.,  eds, The Dense Interstellar Medium in Galaxies. p.~69
  (\mn@eprint {} {astro-ph/0402020})

\bibitem[\protect\citeauthoryear{{Millar} \& {Herbst}}{{Millar} \&
  {Herbst}}{1990}]{Millar}
{Millar} T.~J.,  {Herbst} E.,  1990, \aap, \href
  {http://adsabs.harvard.edu/abs/1990A%26A...231..466M} {231, 466}

\bibitem[\protect\citeauthoryear{{M{\"u}ller}, {Thorwirth}, {Roth}  \&
  {Winnewisser}}{{M{\"u}ller} et~al.}{2001}]{Muller}
{M{\"u}ller} H.~S.~P.,  {Thorwirth} S.,  {Roth} D.~A.,   {Winnewisser} G.,
  2001, \mn@doi [AAP] {10.1051/0004-6361:20010367}, \href
  {http://adsabs.harvard.edu/abs/2001A%26A...370L..49M} {370, L49}

\bibitem[\protect\citeauthoryear{{Neufeld}, {Falgarone}, {Gerin}
  et~al.}{{Neufeld} et~al.}{2012}]{Neufeld}
{Neufeld} D.~A.,  {Falgarone} E.,  {Gerin} M.,   et~al., 2012, \mn@doi [AAP]
  {10.1051/0004-6361/201218870}, \href
  {http://adsabs.harvard.edu/abs/2012A%26A...542L...6N} {542, L6}

\bibitem[\protect\citeauthoryear{{Offringa}, {de Bruyn}, {Biehl}
  et~al.}{{Offringa} et~al.}{2010}]{Offringa10}
{Offringa} A.~R.,  {de Bruyn} A.~G.,  {Biehl} M.,   et~al., 2010, \mn@doi
  [MNRAS] {10.1111/j.1365-2966.2010.16471.x}, \href
  {http://adsabs.harvard.edu/abs/2010MNRAS.405..155O} {405, 155}

\bibitem[\protect\citeauthoryear{{Offringa}, {van de Gronde}  \&
  {Roerdink}}{{Offringa} et~al.}{2012}]{Offringa12}
{Offringa} A.~R.,  {van de Gronde} J.~J.,   {Roerdink} J.~B.~T.~M.,  2012,
  \mn@doi [AAP] {10.1051/0004-6361/201118497}, \href
  {http://adsabs.harvard.edu/abs/2012A%26A...539A..95O} {539, A95}

\bibitem[\protect\citeauthoryear{{Offringa}, {McKinley}, {Hurley-Walker}
  et~al.}{{Offringa} et~al.}{2014}]{Offringa14}
{Offringa} A.~R.,  {McKinley} B.,  {Hurley-Walker} N.,   et~al., 2014, \mn@doi
  [MNRAS] {10.1093/mnras/stu1368}, \href
  {http://adsabs.harvard.edu/abs/2014MNRAS.444..606O} {444, 606}

\bibitem[\protect\citeauthoryear{{Offringa}, {Wayth}, {Hurley-Walker}
  et~al.}{{Offringa} et~al.}{2015a}]{Offringa15}
{Offringa} A.~R.,  {Wayth} R.~B.,  {Hurley-Walker} N.,   et~al., 2015a, \mn@doi
  [PASA] {10.1017/pasa.2015.7}, \href
  {http://adsabs.harvard.edu/abs/2015PASA...32....8O} {32, 8}

\bibitem[\protect\citeauthoryear{{Offringa}, {Wayth}, {Hurley-Walker},
  {Kaplan}, {Barry}  et~al.}{{Offringa} et~al.}{2015b}]{OffriingaRFI}
{Offringa} A.~R.,  {Wayth} R.~B.,  {Hurley-Walker} N.,  {Kaplan} D.~L.,
  {Barry} N.,   et~al., 2015b, \mn@doi [\pasa] {10.1017/pasa.2015.7}, \href
  {http://adsabs.harvard.edu/abs/2015PASA...32....8O} {32, e008}

\bibitem[\protect\citeauthoryear{{Offringa}, {Trott}, {Hurley-Walker}
  et~al.}{{Offringa} et~al.}{2016}]{Offringa16}
{Offringa} A.~R.,  {Trott} C.~M.,  {Hurley-Walker} N.,   et~al., 2016, \mn@doi
  [MNRAS] {10.1093/mnras/stw310}, \href
  {http://adsabs.harvard.edu/abs/2016MNRAS.458.1057O} {458, 1057}

\bibitem[\protect\citeauthoryear{{Olmi}, {Araya}, {Hofner}, {Molinari}
  et~al.}{{Olmi} et~al.}{2014}]{Olmi2}
{Olmi} L.,  {Araya} E.~D.,  {Hofner} P.,  {Molinari} S.,   et~al., 2014,
  \mn@doi [AAP] {10.1051/0004-6361/201322978}, \href
  {http://adsabs.harvard.edu/abs/2014A%26A...566A..18O} {566, A18}

\bibitem[\protect\citeauthoryear{{Olmi}, {Persson}  \& {Codella}}{{Olmi}
  et~al.}{2015}]{Olmi}
{Olmi} L.,  {Persson} C.~M.,   {Codella} C.,  2015, \mn@doi [AAP]
  {10.1051/0004-6361/201526901}, \href
  {http://adsabs.harvard.edu/abs/2015A%26A...583A.125O} {583, A125}

\bibitem[\protect\citeauthoryear{{Olofsson}, {Johansson}, {Hjalmarson}  \&
  {Nguyen-Quang-Rieu}}{{Olofsson} et~al.}{1982}]{Olofsson}
{Olofsson} H.,  {Johansson} L.~E.~B.,  {Hjalmarson} A.,   {Nguyen-Quang-Rieu}
  1982, AAP, \href {http://adsabs.harvard.edu/abs/1982A%26A...107..128O} {107,
  128}

\bibitem[\protect\citeauthoryear{{Perley}, {Chandler}, {Butler}  \&
  {Wrobel}}{{Perley} et~al.}{2011}]{JVLA}
{Perley} R.~A.,  {Chandler} C.~J.,  {Butler} B.~J.,   {Wrobel} J.~M.,  2011,
  \mn@doi [APJL] {10.1088/2041-8205/739/1/L1}, \href
  {http://adsabs.harvard.edu/abs/2011ApJ...739L...1P} {739, L1}

\bibitem[\protect\citeauthoryear{{Pickett}, {Poynter}, {Cohen}, {Delitsky},
  {Pearson}  \& {M{\"u}ller}}{{Pickett} et~al.}{1998}]{Pickett}
{Pickett} H.~M.,  {Poynter} R.~L.,  {Cohen} E.~A.,  {Delitsky} M.~L.,
  {Pearson} J.~C.,   {M{\"u}ller} H.~S.~P.,  1998, \mn@doi [JQSRT]
  {10.1016/S0022-4073(98)00091-0}, \href
  {http://adsabs.harvard.edu/abs/1998JQSRT..60..883P} {60, 883}

\bibitem[\protect\citeauthoryear{{Quintana-Lacaci}, {Ag{\'u}ndez}, {Cernicharo}
   et~al.}{{Quintana-Lacaci} et~al.}{2013}]{Quintana-Lacaci}
{Quintana-Lacaci} G.,  {Ag{\'u}ndez} M.,  {Cernicharo} J.,   et~al., 2013,
  \mn@doi [AAP] {10.1051/0004-6361/201322728}, \href
  {http://adsabs.harvard.edu/abs/2013A%26A...560L...2Q} {560, L2}

\bibitem[\protect\citeauthoryear{{Remazeilles}, {Dickinson}, {Banday},
  {Bigot-Sazy}  \& {Ghosh}}{{Remazeilles} et~al.}{2015}]{Remazeilles}
{Remazeilles} M.,  {Dickinson} C.,  {Banday} A.~J.,  {Bigot-Sazy} M.-A.,
  {Ghosh} T.,  2015, \mn@doi [MNRAS] {10.1093/mnras/stv1274}, \href
  {http://adsabs.harvard.edu/abs/2015MNRAS.451.4311R} {451, 4311}

\bibitem[\protect\citeauthoryear{{Rudnitskij}}{{Rudnitskij}}{2002}]{Rudnitskij02}
{Rudnitskij} G.~M.,  2002, \mn@doi [PASA] {10.1071/AS02018}, \href
  {http://adsabs.harvard.edu/abs/2002PASA...19..499R} {19, 499}

\bibitem[\protect\citeauthoryear{{Sault}, {Teuben}  \& {Wright}}{{Sault}
  et~al.}{1995}]{Sault}
{Sault} R.~J.,  {Teuben} P.~J.,   {Wright} M.~C.~H.,  1995, in {Shaw} R.~A.,
  {Payne} H.~E.,   {Hayes} J.~J.~E.,  eds,  Astronomical Society of the Pacific
  Conference Series Vol. 77, Astronomical Data Analysis Software and Systems
  IV. p.~433 (\mn@eprint {} {astro-ph/0612759})

\bibitem[\protect\citeauthoryear{{Sevenster}, {van Langevelde}, {Moody}
  et~al.}{{Sevenster} et~al.}{2001}]{Sevenster}
{Sevenster} M.~N.,  {van Langevelde} H.~J.,  {Moody} R.~A.,   et~al., 2001,
  \mn@doi [AAP] {10.1051/0004-6361:20000354}, \href
  {http://adsabs.harvard.edu/abs/2001A%26A...366..481S} {366, 481}

\bibitem[\protect\citeauthoryear{{Soszy{\'n}ski} et~al.,}{{Soszy{\'n}ski}
  et~al.}{2013}]{Soszynski13}
{Soszy{\'n}ski} I.,  et~al., 2013, ACTAA, \href
  {http://adsabs.harvard.edu/abs/2013AcA....63...21S} {63, 21}

\bibitem[\protect\citeauthoryear{{Spitzer}}{{Spitzer}}{1998}]{SpitzerISM}
{Spitzer} L.,  1998, {Physical Processes in the Interstellar Medium}

\bibitem[\protect\citeauthoryear{{Taylor} et~al.,}{{Taylor}
  et~al.}{2012}]{Taylor}
{Taylor} G.~B.,  et~al., 2012, \mn@doi [JAI] {10.1142/S2251171712500043}, \href
  {http://adsabs.harvard.edu/abs/2012JAI.....150004T} {1, 50004}

\bibitem[\protect\citeauthoryear{{Tercero}, {Cernicharo}, {L{\'o}pez}
  et~al.}{{Tercero} et~al.}{2015}]{Tercero15}
{Tercero} B.,  {Cernicharo} J.,  {L{\'o}pez} A.,   et~al., 2015, \mn@doi [AAP]
  {10.1051/0004-6361/201526255}, \href
  {http://adsabs.harvard.edu/abs/2015A%26A...582L...1T} {582, L1}

\bibitem[\protect\citeauthoryear{{Thiagaraj}, {Srivani}, {Roshi}
  et~al.}{{Thiagaraj} et~al.}{2015}]{Thiagaraj}
{Thiagaraj} P.,  {Srivani} K.~S.,  {Roshi} D.~A.,   et~al., 2015, \mn@doi [EA]
  {10.1007/s10686-015-9444-3}, \href
  {http://adsabs.harvard.edu/abs/2015ExA....39...73P} {39, 73}

\bibitem[\protect\citeauthoryear{{Tingay}, {Goeke}, {Bowman}  et~al.}{{Tingay}
  et~al.}{2013}]{Tingay13}
{Tingay} S.~J.,  {Goeke} R.,  {Bowman} J.~D.,   et~al., 2013, \mn@doi [PASA]
  {10.1017/pasa.2012.007}, \href
  {http://adsabs.harvard.edu/abs/2013PASA...30....7T} {30, 7}

\bibitem[\protect\citeauthoryear{{Walsh}, {Breen}, {Britton}, {Brooks},
  {Burton}, {Cunningham}  \& {et al}}{{Walsh} et~al.}{2011}]{Walsh11}
{Walsh} A.~J.,  {Breen} S.~L.,  {Britton} T.,  {Brooks} K.~J.,  {Burton} M.~G.,
   {Cunningham} M.~R.,   {et al} 2011, \mn@doi [MNRAS]
  {10.1111/j.1365-2966.2011.19115.x}, \href
  {http://adsabs.harvard.edu/abs/2011MNRAS.416.1764W} {416, 1764}

\bibitem[\protect\citeauthoryear{{Wilson} \& {Rood}}{{Wilson} \&
  {Rood}}{1994}]{Wilson}
{Wilson} T.~L.,  {Rood} R.,  1994, \mn@doi [ARAA]
  {10.1146/annurev.aa.32.090194.001203}, \href
  {http://adsabs.harvard.edu/abs/1994ARA%26A..32..191W} {32, 191}

\bibitem[\protect\citeauthoryear{{Yamamura}, {Kawaguchi}  \&
  {Ridgway}}{{Yamamura} et~al.}{2000}]{YK2000}
{Yamamura} I.,  {Kawaguchi} K.,   {Ridgway} S.~T.,  2000, \mn@doi [APJL]
  {10.1086/312420}, \href {http://adsabs.harvard.edu/abs/2000ApJ...528L..33Y}
  {528, L33}

\bibitem[\protect\citeauthoryear{{Zheng}, {Tegmark}, {Dillon}, {Kim}, {Liu}
  et~al.}{{Zheng} et~al.}{2017}]{Zheng}
{Zheng} H.,  {Tegmark} M.,  {Dillon} J.~S.,  {Kim} D.~A.,  {Liu} A.,   et~al.,
  2017, \mn@doi [MNRAS] {10.1093/mnras/stw2525}, \href
  {http://adsabs.harvard.edu/abs/2017MNRAS.464.3486Z} {464, 3486}

\bibitem[\protect\citeauthoryear{{van Dishoeck}}{{van
  Dishoeck}}{2014}]{2014FaDi..168....9V}
{van Dishoeck} E.~F.,  2014, \mn@doi [Faraday Discussions]
  {10.1039/C4FD00140K}, \href
  {http://adsabs.harvard.edu/abs/2014FaDi..168....9V} {168, 9}

\bibitem[\protect\citeauthoryear{{van Haarlem} et~al.,}{{van Haarlem}
  et~al.}{2013}]{vanHaarlem}
{van Haarlem} M.~P.,  et~al., 2013, \mn@doi [AAP]
  {10.1051/0004-6361/201220873}, \href
  {http://adsabs.harvard.edu/abs/2013A%26A...556A...2V} {556, A2}

\makeatother
\end{thebibliography}

%%%%%%%%%%%%%%%%%%%%%%%%%%%%%%%%%%%%%%%%%%%%%%%%%%

%%%%%%%%%%%%%%%%% APPENDICES %%%%%%%%%%%%%%%%%%%%%

\appendix

 \section{\\Column Density } \label{Column}
Column density is a quantity used in understanding the physical and chemical nature of an object, as well as a starting point for analysis of isomer ratios, isotopomer ratios, and chemical composition \citep{Mangum}.  The calculation of the column density in the energy at the upper excitation level level (N$_{\mathrm{u}}$), when the optical depth is assumed to be small and the lines are in emission, is given by:

\begin{equation}
N_{u} = \frac{8\mathrm{k}\pi\nu^{2}}{A_{\mathrm{ul}}\mathrm{h}\mathrm{c}^{3}}  \int T_{\mathrm{b}}\, \mathrm{d}v
\end{equation} 

where $\nu$ is the rest frequency, k is Boltzmannn's Constant,  h is Planck's constant, c is the speed of light, and $ \int {T_{\mathrm{b}}}\, \mathrm{dv}$ represents the integrated intensity of the spectral line along the velocity axis (v).  The Einstein coefficient for the transition, A$_{\mathrm{ul}}$, is calculated using the formula: 

\begin{equation}
A_{\mathrm{ul}} = \frac{16\pi\nu^{3}}{3\varepsilon_{o}{\mathrm{hc}}^{3}}\mu^{2}
\end{equation}

where $\varepsilon_{o}$ is the permittivity of free space and $\mu$ is the electric dipole moment for a pair of opposite charges.  For molecules that have multiple dimensional components, the vector sum is calculated by taking the square root of the summation of the squares of each dimensional value.  

The total column density (N$_{\mathrm{tot}}$) can then be calculated by:

\begin{equation}
N_{\mathrm{tot}} = \frac{N_{\mathrm{u}}}{g_{\mathrm{u}}}e^{\frac{E_{\mathrm{u}}}{kT}}Q_{\mathrm{rot}}(\mathrm{T}_{\mathrm{ex}})
\end{equation} 

where $E_{\mathrm{u}}$ is the energy in the upper excitation state and $g_{\mathrm{u}}$ is the total degeneracy term for the energy of a particular transition.    We also make the assumption that the observed molecule has a single excitation temperature, T$_{\mathrm{ex}}$=T$_{\mathrm{rot}}$ \citep{Hildebrand}.  The $Q_{\mathrm{rot}}$ is the rotational partition function at an assumed temperature, for which we use 9.375\,K based on the analysis done be \cite{Jones13} that showed that T$_{\mathrm{ex}}$ of 10\,K is reasonable for the central molecular zone. However, assuming a excitation temperature of 18\,K would increase the total column density by approximately 12.  The $Q_{\mathrm{rot}}$, $E_{\mathrm{u}}$ and $g_{\mathrm{u}}$ values are taken from the CDMS.

When the transitions are in absorption, a correction to the total column density must be made to account for the strong continuum.  As described by \cite{Comito}, the total column density ($N_{\mathrm{tot}}$ ) then becomes:

\begin{equation}
N_{\mathrm{tot}} = \frac{8 \pi\nu^{3}}{A_{\mathrm{ul}}{\mathrm{c}}^{3}}\frac{g_{\mathrm{l}}}{g_{\mathrm{u}}} \tau \Delta v
\end{equation} 

where $\Delta$$v$ is the change in velocity across the transition and $\tau$ is the optical depth defined as:

\begin{equation}
\tau = -{\mathrm{ln}}[1-\frac{T_{{\mathrm{L}}}}{T_{{\mathrm{C}}}}]
\end{equation}

The value of $T_{{\mathrm{C}}}$ is the brightness temperature of the continuum which is derived from scaling the Haslam 408\,MHz \citep{Remazeilles} map to the frequency of the molecular line, using a spectral index of $-$2.5 \citep{Zheng}.  $T_{{\mathrm{L}}}$ is the brightness temperature of the spectral line, using the Raleigh-Jeans approximation.  All values and constants use standard (SI) units.

A standard assumption in the calculation of column density is that the source fills the telescope synthesised beam \citep{Herbst09}. When this assumption is not valid, beam dilution occurs, and a correction must be made.  The FWHM of the synthesised beam for the MWA at 120\,MHz is three arc minutes and the objects of interest may be much smaller. The column densities can be scaled to the assumed size of the source by a ratio of solid angles.

With a channel resolution of 26\,km\,s$^{-1}$, it is expected that most lines will emit in a single channel.  However, the area under the curve used in determining the integrated intensity will be preserved when compared to higher spectral resolution instruments in which emission will span multiple channels.  Therefore, no correction is required for channel smearing.  

%%%%%%%%%%%%%%%%%%%%%%%%%%%%%%%%%%%%%%%%%%%%%%%%%%

% Don't change these lines
\bsp	% typesetting comment
\label{lastpage}
\end{document}